\documentclass[twocolumn, preprintnumbers, showpacs, nofootinbib, aps]{revtex4}
\usepackage{graphicx}
\usepackage{amssymb}
\usepackage{amsmath}
\usepackage{epstopdf}
\usepackage{color}
\usepackage{mathrsfs}
\usepackage{subfig}
\begin{document}

\def\SU2{\ensuremath{{SU(2)}}}
\def\sdmass{\ensuremath{\sqrt{a_o}/2}}
\newcommand{\bea}{\begin{eqnarray}}
\newcommand{\eea}{\end{eqnarray}}
\def\half{\frac{1}{2}}
\def\beq{\begin{equation}}
\def\eeq{\end{equation}}
\def\beqn{\begin{eqnarray}}
\def\eeqn{\end{eqnarray}}

 \title{Mass Inflation in the Loop Black Hole}
\author{Eric G. Brown, Robert Mann}
 \affiliation{Department of Physics and Astronomy, University of Waterloo, Waterloo, Ontario N2L 3G1, Canada}
\author{Leonardo Modesto}
\affiliation{Perimeter Institute for Theoretical Physics, 31 Caroline St.N., Waterloo, ON N2L 2Y5, Canada }

%\ead{lmodesto@perimeterinstitute.ca}

\begin{abstract}

In classical general relativity the Cauchy horizon within a two-horizon black hole is unstable via a phenomenon known as mass inflation, in which the mass parameter (and the spacetime curvature) of the black hole diverges at the Cauchy horizon.
Here we study this effect for loop black holes -- quantum gravitationally corrected black holes from loop quantum gravity -- whose construction alleviates the $r=0$ singularity present in their classical counterparts.   We use a simplified model of mass inflation, which makes use of the generalized DTR relation, to conclude that the Cauchy horizon of loop black holes indeed results in a curvature singularity similar to that found in classical black holes. The DTR relation is of particular utility in the loop black hole because it does not directly rely upon Einstein's field equations.  We elucidate some of the interesting and counterintuitive properties of the loop black hole, and corroborate our results using an alternate
model of mass inflation due to  Ori.
\end{abstract}

\maketitle

\tableofcontents

\section{Introduction}
%motivation and explanation of the problem

A significant hinderance to the development of the correct theory of quantum gravity lies in the fact that, by its very nature, any predicted phenomenology is likely to be beyond the reach of practical experimental capability. Without any experimental guidance (or at least very little) it seems that we are merely left to postulate how quantum gravity may alter classical theory in high energy regimes, with constraints on the theory coming in the form of self-consistency checks.  Despite this limitation there are still predictions that can be used to guide us in our quest to find the fundamental theory. Of particular interest in this respect is how quantum gravity may change the presence of the singularities that are predicted by general relativity. It has long been expected that the infinities that occur, for example inside black holes, represent a breakdown of Einstein's theory and that a proper quantum gravitational treatment   will be able to placate them. This then provides us with a ``laboratory" that can be used to test the merit of a proposed theory of quantum gravity; for example a theory predicting that the interiors of black holes are regular can be regarded as an improvement over those that do not.
 %This said, one should be careful not to let the desired prediction dictate too strongly the physics performed since in this case the prediction becomes not a prediction at all but rather a necessary conclusion. This is not to say that the expectation of a singularity-free spacetime can not motivate the formulation of a quantum gravity theory, but just that top-down tinkering can not be used when the goal is to test a pre-existing theory.

With this as motivation, much work has been performed to say how black holes might be modified in the context of quantum gravity. Several proposals from several different approaches to quantum gravity have been given that are free of the classical \(r=0\) singularity \cite{Modesto:2008im,bh2,bh3,bh4,bh5}. Despite being regular at \(r=0\) however, these quantum  black holes generically have more than one horizon. One therefore ought to be wary of their overall regularity since it is well known that classical black holes with more than one horizon also exhibit divergent behaviour at their inner horizon (the Cauchy horizon).  A demonstration that  quantum black holes also exhibit divergent behaviour at their Cauchy horizon  will indicate that they are not performing the job that is expected of them; i.e. they do not represent regular black holes.  At the least this will imply some insufficiency in
the assumptions and (semiclassical) approximations used in constructing the solutions, if not in the underlying
  theory of quantum gravity from whence they came.

In classical gravity, the instability of the Cauchy horizon can be seen to arise from the infinite blueshift of external radiation that occurs on the inner horizon. This phenomenon is most easily demonstrated in the Reissner-Nordstr\(\ddot{\text{o}}\)m black hole, but can also be shown to arise in the Kerr and Kerr-Newmann black holes. This type of instability had been thought to occur for many years and was demonstrated to first order by observing the behaviour of perturbations on the metric \cite{RNstability,RNstability2}. Such an analysis, however, says nothing about the backreaction on the spacetime that is induced. It was not until 1990 that Poisson and Israel showed rigorously that in a realistic black hole the Cauchy horizon of a Reissner-Nordstr\(\ddot{\text{o}}\)m spacetime is unstable via a phenomenon that they dubbed \emph{mass inflation}.  The Cauchy horizon generically develops singular curvature  caused by the local mass parameter that grows indefinitely large \cite{poisson}.

In this paper we ask whether or not mass inflation occurs in a class of quantum gravitational black holes motivated by loop quantum gravity. We will see that a mass-inflation-like phenomenon indeed seems to occur. Although we work within the context of a specific model, our results depend on quite general features of the black hole interior, suggesting
that any black hole with a Cauchy horizon -- classical or quantum -- will be hard pressed to avoid the same result.

Loop quantum gravity  (LQG) \cite{LQGgeneral,LQGgeneral2,LQGgeneral3}, has given rise to
models that afford a description of  the very early universe. This simplified
framework, which uses a minisuperspace approximation, has been shown
to resolve the initial singularity problem \cite{Bojowald}. A black hole metric in this model, known as the loop black hole (LBH)
\cite{Modesto:2008im}, was constructed from a modification to the holonomic version of the Hamiltonian constraint.  A parametric function $\sigma(\delta)$, that labels the elements in the class of Hamiltonian constraints compatible with spherical symmetry and homogeneity, can be uniquely fixed from asymptotic boundary conditions, yielding the proper classical Hamiltonian in the limit the polymeric parameter \(\delta \rightarrow 0\).

As such, this black hole solution might be expected to
model features induced by LQG effects. It has  a property of self-duality
that removes the \(r=0\) singularity and replaces it with
another asymptotically flat region. Both the
thermodynamic properties \cite{Modesto:2008im,poly} and the dynamical aspects of collapse and
evaporation \cite{Hossenfelder:2009fc} of these self-dual black holes have been previously studied.
However, since the polymerization of  the Hamiltonian constraint in the homogeneous region is
inside the event horizon, the physical meaning of the solution
when the metric is analytically continued to all spacetime remains an open problem.
Nevertheless this metric is useful insofar as it can be expected to provide a
first approximation to black hole solutions that emerge from LQG. Since its interior has a Cauchy horizon, we
can investigate its stability in this quantum gravitational framework.
Black hole spacetimes have also been investigated in a midi-superspace reduction of LQG \cite{GP}.
While the black hole solution obtained by this method is closer to the full LQG theory, it can only be presented in numerical
form, whereas the black hole we consider has a closed analytic form and so can be more
easily investigated. We emphasize that our analysis has limitations insofar as it does not correspond to a full LQG solution.

Our paper is organized as follows. In Sect. \ref{loopderiv} we
recall the loop black hole (LBH) derivation in short and expound some of its properties. In Sect. \ref{preliminary} we consider a short perturbative calculation that indicates the instability of the Cauchy horizon and motivates further analysis. In Sect. \ref{DTR} we will review the derivation of the so called \emph{generalized DTR relation}. This result will play a vital role in Sect. \ref{massinflation}, where we will see how mass inflation arises in the loop black hole before concluding in Sect. \ref{conclusion}.  We consider in
an Appendix a model of mass inflation developed by Ori \cite{ori} applied to the LBH.  Notwithstanding
certain difficulties entailed in this approach, we find that it corroborates the main results in our paper.

\section{The Loop Black Hole}
\label{loopderiv}

The regular black hole metric that we will be using is derived from a simplified model of LQG \cite{Modesto:2008im}.
LQG is based on a canonical quantization of the Einstein equations written in terms of the Ashtekar variables \cite{AA}, that is in terms of an $SU(2)$ 3-dimensional connection $A$ and a triad $E$. The basis states of LQG then are closed graphs, the edges of which are labeled by irreducible $SU(2)$ representations and the vertices by $SU(2)$ intertwiners (for a review see e.g. \cite{LQGgeneral,LQGgeneral2,LQGgeneral3}). The
edges of the graph represent quanta of area with area $\gamma l_P^2 \sqrt{j(j+1)}$, where $j$ is a half-integer representation label on the edge, $l_P$ is the Planck length,  and $\gamma$ is a parameter of order $1$ called the Immirzi parameter. The vertices of the graph represent quanta of $3$-volume. One important consequence that we will use in the following is that the area is quantized and the smallest possible quanta correspond to an area of $\sqrt{3}/2 \gamma l_P^2$.

To obtain the simplified black hole model the following assumptions were made. First, the number of variables was reduced by assuming spherical symmetry. Then, instead of all possible closed graphs, a regular lattice with edge lengths $\delta_b$ and $\delta_c$ was used. The solution was then obtained dynamically inside the homogeneous region (that is inside the horizon where space is homogeneous but not static, a spacetime known as Kantowski-Sachs spacetime). An analytic continuation to the outside of the horizon shows that one can reduce the two free parameters by identifying the minimum area present in the solution with the minimum area of LQG. The one remaining unknown constant $\delta_b$  is a
dimensionless parameter of the model that determines the strength of deviations from the classical theory, and would have to be constrained by experiment. Redefining  $\delta_b = \delta$, the free parameter that appears in the metric is
$\epsilon = \delta\gamma$, the product of the Immirzi parameter $\gamma$  and the polymeric quantity $\delta$.
With the plausible expectation that   quantum gravitational corrections become relevant only when the curvature is in the Planckian regime, corresponding to $\delta \gamma  < 1$, outside the horizon the solution is the Schwarzschild solution up to negligible Planck-scale corrections in $l_P$ and $\delta \gamma$.

The procedure to obtain the metric is  the following.
\begin{enumerate}
\renewcommand{\theenumi}{\roman{enumi}}
\renewcommand{\labelenumi}{\theenumi}
\item We define the Hamiltonian constraint by replacing the homogeneous connection with
the holonomies along the fixed graph decided above. The diff-constraint is identically zero
because of homogeneity and the Gauss constraint is zero for the Kantowski-Sachs spacetime.
\item We solve the Hamilton equation of motion for the holonomic Hamiltonian system
imposing the Hamiltonian constraint to be zero. % at any instant.
\item The third step consists in extending the solution to all spacetime. Note that while there is no
mathematical obstruction to this extension it is an assumption, since we cannot know  the correct polymerization of the Hamiltonian constraint in the full spacetime  as we found the solution in the homogeneous region. 
%However the promising features of the LBH provide confidence that such a polymerization exists.
\end{enumerate}
A more technical point is the following.
In the homogeneous region a cut-off in the spatial direction is introduced. This extra structure disappears in the metric,  and all physical quantities are  independent of this length scale  \cite{Modesto:2008im}.
%\end{enumerate}

The above procedure then yields a black hole solution that can reasonably be expected to
model features induced by LQG effects.
This quantum gravitationally corrected
Schwarzschild metric -- known as the Loop Black Hole (LBH) -- can be expressed in the form
\begin{eqnarray}
ds^2 = - G(r) dt^2 + \frac{dr^2}{F(r)} + H(r) d\Omega^2,
\label{g}
\end{eqnarray}
with $d \Omega^2 = d \theta^2 + \sin^2 \theta d \phi^2$ and
\begin{eqnarray}
&& G(r) = \frac{(r-r_+)(r-r_-)(r+ r_{*})^2}{r^4 +a_o^2}~ , \nonumber \\
&& F(r) = \frac{(r-r_+)(r-r_-) r^4}{(r+ r_{*})^2 (r^4 +a_o^2)} ~, \nonumber \\
&& H(r) = r^2 + \frac{a_o^2}{r^2}~ .
\label{statgmunu}
\end{eqnarray}
Here, $r_+ = 2m$ and $r_-= 2 m P^2$ are the two horizons, and \(r_* \equiv \sqrt{r_+ r_-} = 2mP\). $P$ is the
polymeric parameter \(P \equiv (\sqrt{1+\epsilon^2} -1)/(\sqrt{1+\epsilon^2} +1)\), with
$\epsilon = \delta\gamma \ll 1$. Hence $P \ll 1$,  implying $r_-$ and $r_*$ are very close to $r=0$. The area $a_o$ is equal to $A_{\rm min}/8 \pi$, $A_{\rm min}$ being the minimum area gap of LQG.
%With $a_o=A_{\rm Min}/8 \pi$, it is of the order of the Planck length squared.

Note that in the above metric, $r$ is only asymptotically the usual radial
coordinate since $g_{\theta \theta}$ is not just $r^2$.  This choice of
coordinates  has the advantage of easily revealing the properties
of this metric. Most importantly, in the limit
$r \to \infty$, the deviations from the Schwarzschild-solution are of
order $M \epsilon^2/r$, where $M$ is the usual ADM-mass:
\beqn
G(r) &\to& 1-\frac{2 M}{r} (1 - \epsilon^2)~, \nonumber  \\
F(r) &\to& 1-\frac{2 M}{r}~ , \nonumber \\
H(r) &\to& r^2 .
\eeqn
The ADM mass is the mass inferred by an observer at flat asymptotic infinity; it is determined solely
by the metric at asymptotic infinity.  The parameter $m$ in the solution is related to the mass $M$ by $M = m (1+P)^2$.
%and $\epsilon = 4 P\ll1$.

Performing the coordinate transformation $R = a_o/r$ with the rescaling
$\tilde t= t \, r_*^{2}/a_o$, and
simultaneous  replacements $R_\pm = a_o/r_\mp$, $R_* = a_o/r_*$ yields the result that the metric in
the new coordinates has the same form as that in the old coordinates, thus exhibiting a
very compelling type of self-duality with dual radius $r=\sqrt{a_o}$. Looking at the angular part
of the metric, one sees that this dual radius corresponds to a minimal possible
surface element. It is then also clear that in the limit $r\to 0$, corresponding
to $R\to \infty$, the solution
does not have a singularity, but instead has another asymptotically flat Schwarzschild region.

Computing the surface gravity
\begin{eqnarray}
\kappa^2 = - g^{\mu \nu} g_{\rho \sigma} \nabla_{\mu} \chi^{\rho} \nabla_{\nu}
\chi^{\sigma}/2 = - g^{\mu \nu} g_{\rho \sigma}  \Gamma^{\rho}_{\;\mu 0} \Gamma^{\sigma}_{\;\nu 0}/2,
\end{eqnarray}
where $\chi^{\mu}=(1,0,0,0)$ is a timelike Killing vector
in $r>r_+$ and $r<r_-$ (but spacelike
in $r_- <r < r_+$)   yields
\begin{eqnarray} \label{surfgrav}
 \kappa_- =  \frac{4 m^3 P^4 (1-P^2)}{16 m^4 P^8 + a_o^2}, \,\,\,\,\,\,
 \kappa_+ = \frac{4 m^3 (1-P^2)}{16 m^4 + a_o^2}.
\label{kpm}
\end{eqnarray}
for the surface gravity on the
inner and outer horizons of  the metric (\ref{g}).

We now go on to a simple calculation that both introduces some of the background structure necessary for investigating
mass inflation and gives a first illustration of the troublesome behaviour exhibited by the Cauchy horizon.

\section{A Preliminary Calculation of Stability}
\label{preliminary}

% change form of luminosity
% explain limitations of this calculation (no backreaction)
% show finite curvature when backreaction is considered (note that resulting stress-energy is NOT null dust)

%also, cite your paper showing better stability properties

In this section we are going to perform a `warm up' calculation  for mass inflation.  We start with the LBH given by Eq. (\ref{g}), though the same calculation is easily performed for any spherically symmetric black hole.
Consider a stream of null dust entering the black hole. We neglect the backreaction that this radiation has on the spacetime and we compute the energy density that a timelike observer would measure from it as they cross the Cauchy horizon.

\begin{figure}[h]
 % \caption{A Penrose diagram displaying the spacetime outside of the black hole as well as between the horizons. We illustrate the ingoing null radiation and label the direction of increasing advanced time \(v\).}
 \centering
  \includegraphics[width=0.45\textwidth]{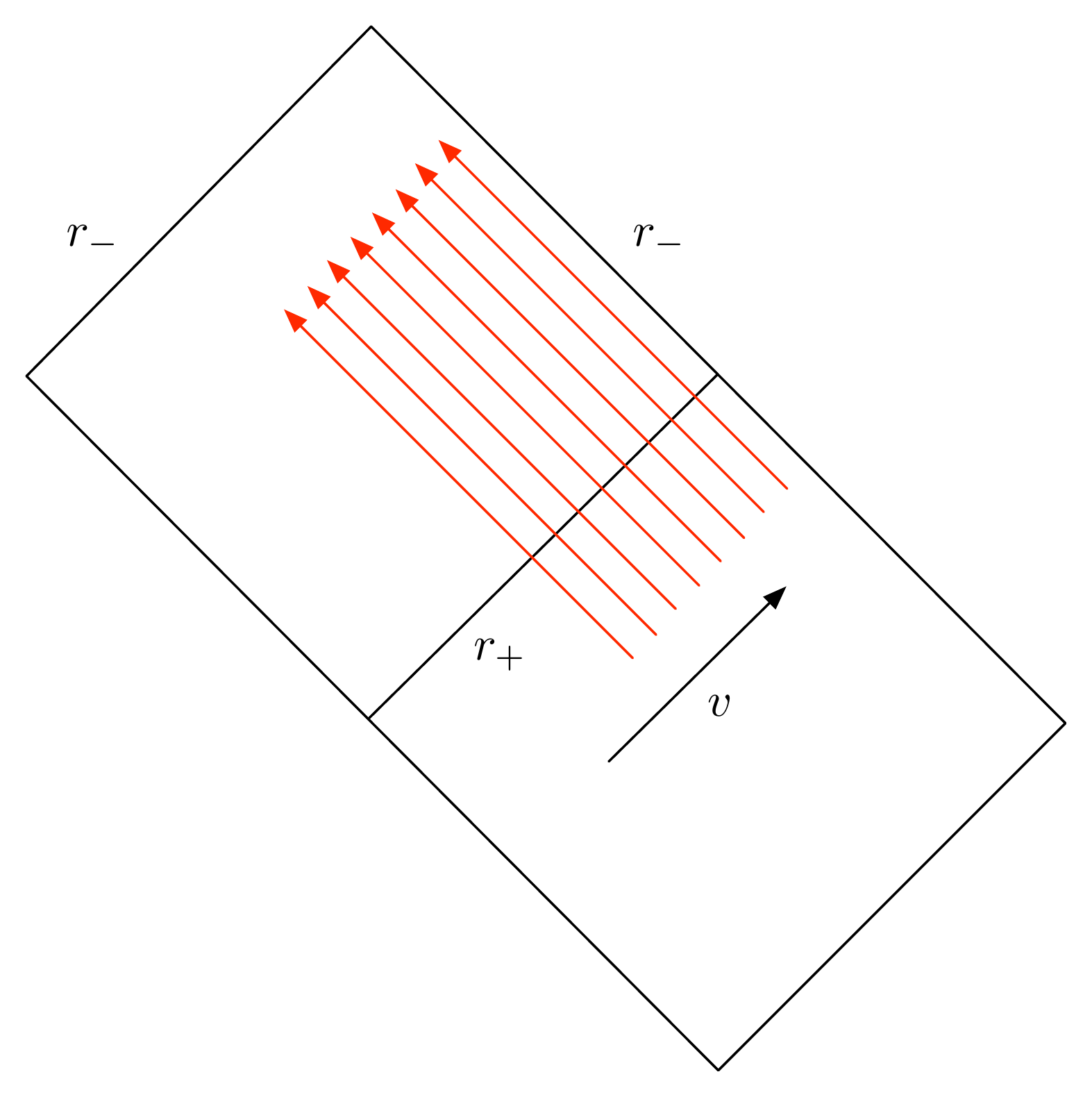}
    \caption{A Penrose diagram displaying the spacetime outside of the black hole as well as between the horizons. We illustrate the ingoing null radiation and label the direction of increasing advanced time \(v\).}
\label{penrose0}
  \end{figure}

We will find it useful to define the advanced time \(v \equiv t+r^*\), where \(r^*\) is the tortoise coordinate (not to be confused with the variable \(r_* \equiv 2mP\) present in the metric) defined by
\begin{align}
    \frac{dr^*}{dr} \equiv \frac{1}{\sqrt{G(r)F(r)}} .
\end{align}
The limit \(v=\infty\) corresponds to future null infinity when outside of the black hole and to the Cauchy horizon when inside. Thus, when we speak of ``approaching the Cauchy horizon" we are referring to the \(v \rightarrow \infty\) limit.  It has been shown \cite{price} that  for a black hole formed by gravitational collapse there will always be ingoing null radiation with energy density falling off as an inverse power, \(\sim v^{-\gamma}\) where \(\gamma \geq 12\), for large \(v\). In any gravitational collapse there will always be gravitational radiation that escapes into space, but some of this radiation will later be scattered back by the gravitational potential present in the spacetime surrounding the black hole. The inverse power law given here corresponds to the luminosity of this backscattered radiation, and its importance lies in the fact that it is expected to always be present in a realistic black hole.  We
make use of this result for the LBH,  since the gravitational emission and backscatter in question occurs outside of the event horizon, where the LBH spacetime only negligibly deviates from that of a classical Schwarzschild black hole, for which Price's result is known to apply. This situation is depicted in Fig. \ref{penrose0}.

First we cast the loop metric in \((v,r,\theta,\phi)\) coordinates, obtaining
\begin{align}
    ds^2=-G(r)dv^2+2\sqrt{\frac{G(r)}{F(r)}}drdv+H(r)d\Omega^2 ,
\end{align}
where the metric functions are as defined in Sect. \ref{loopderiv}.
We then consider a stress-energy tensor corresponding to   ingoing null dust. It takes the form
\begin{align}
    T_{\alpha \beta}=\mu(r,v)(\partial_\alpha v) (\partial_\beta v) .
\end{align}

With these last two equations we determine the form of \(\mu(r,v)\) by computing the conservation condition \(0=T^{\alpha \beta}_{\;\;\; ; \beta}\). The result is that we must have \(\mu(r,v)=A(r)L(v)\), where \(L(v)\) is an arbitrary function of \(v\) and \(A(r)\) must satisfy the equation
\begin{align}
    0=r(r^4+a_0^2)\frac{dA(r)}{dr}+2(r^4-a_0^2)A(r), \nonumber
\end{align}
the solution of which is
\begin{align}
    A(r) \propto \frac{r^2}{r^4+a_0^2} \nonumber
\end{align}
and so we find
\begin{align}
    \mu(r,v)=\frac{r^2}{4\pi(r^4+a_0^2)}L(v)
\end{align}
for the general form of $\mu(r,v)$.

Far from the black hole the spacetime is well approximated by a Schwarzschild solution, and indeed for \(r >> \sqrt{a_0}\) this form of \(\mu(r,v)\) reverts to the same form as is known for the Schwarzschild black hole. We thus identify \(L(v)\) with the usual luminosity, and we expect the analysis by Price \cite{price} to hold to a good approximation. That is, we have \(L(v) \sim v^{-\gamma}\) with \(\gamma \geq 12\).

In order to determine the energy density a timelike observer would measure from this dust we also need to specify the four-velocity of the observer. For the case at hand, however, all we need is the \(v\) component, which we find to be
\begin{align}
    U^v=\frac{1}{G}(E-\sqrt{E^2-G}),
\end{align}
where \(E\) is a constant of the motion, the sign of which gives the direction of travel between the horizons. \(E>0\) corresponds to an ingoing (left-moving) observer and \(E<0\) to an outgoing one (right-moving). We want to focus on right-moving observers, since it is they who will reach the right branch of the Cauchy horizon (the branch located at \(v=\infty\)); we thus only consider geodesics with negative \(E\). The energy density measured by such an observer is given by
\begin{align}
    \rho=T_{\alpha \beta}U^\alpha U^\beta=\mu(r,v)(U^v)^2.
\end{align}

As the Cauchy horizon is approached we have \(G \rightarrow 0\), and so the negativity of \(E\) implies that
\begin{align}
    (U^v)^2 \simeq 4E^2/G^2, 
\end{align}
as this approach takes place. However, it is clear that we also need to know in what manner the metric function approaches zero. For this one can compute the tortoise coordinate \(r^*\) directly from its definition and show that near the Cauchy horizon, \(r \simeq r_-\), it is dominated by a term given by
\begin{align}
    r^*\simeq -(2 \kappa_-)^{-1}\log |r-r_-| , \nonumber
\end{align}
where \(\kappa_-\) is the surface gravity of the Cauchy horizon, given by Eq. (\ref{kpm}). If one then computes the \(t\) and \(r\) components of the four-velocity it can be shown that for a right-moving observer \(\frac{dr^*}{dt}=\frac{dr^*}{dr}\frac{U^r}{U^t} \simeq 1\) near the Cauchy horizon. From this we obtain
\begin{eqnarray}
  && \hspace{-0.5cm} -v=-t-r^*\simeq -2r^*+\text{const} \nonumber
   \\
 && \hspace{-0.5cm}  \simeq \kappa_-^{-1}\log|r-r_-|+\text{const} % \,\,\,\, \nonumber \\
    \implies %\;\;\;\;\;\;\;\;
 \,\,\,\,    (r-r_-)^{-1} \propto e^{\kappa_- v} .  \nonumber
\end{eqnarray}
which, since the metric functions \(F\) and \(G\) are proportional to \(r-r_-\), results in the conclusion that near the Cauchy horizon they decay  as
\begin{align}
    G \; \text{and} \; F \propto e^{-\kappa_- v}, \;\;\;\;\; v \rightarrow \infty.
\end{align}

Thus, we finally find that the energy density measured by the observer goes as
\begin{align}
    \rho \sim e^{2\kappa_- v}L(v) , 
\end{align}
as the Cauchy horizon is approached. Since \(L(v)\) follows an inverse power law, we conclude that the observer measures diverging energy density as they cross of the Cauchy horizon.

A simpler scenario, in which a field pulse is fired in to the black hole and then scattered in the interior geometry, yields a different result \cite{us}.  Its energy density diverges   at the Cauchy horizon  in a manner dependent  on the difference of the surface gravities \(\kappa_-\) and \(\kappa_+\).   Provided  \(\kappa_+ \geq \kappa_-\), no divergence occurs,
and  the loop black hole appears to be more stable under this type of radiation than classical black holes. However there does not appear to be any cure  (within the perturbative framework) for the more generic result obtained above using inverse power law decay.

With this result as motivation we  move on to the mass inflation analysis,  computing the behaviour of the spacetime curvature resulting from the infinitely blue-shifted radiation. We employ  a simplified but well-motivated model that makes the problem tractable. In preparation for this we will briefly review the derivation of the generalized DTR relation, a key ingredient  in our
analysis.

\section{The Generalized DTR Relation}
\label{DTR}

% explain motivation for using DTR. same essential physics.
% cite the two papers
% work with spherical symmetry, but note that end result is independent of this

One of the key ingredients needed for mass inflation to occur is the existence of outgoing null radiation within the black hole, in addition to the ingoing radiation (that which is undergoing infinite blueshift). It is the cross flow between these two streams that    induces the phenomenon. It is expected that in a realistic black hole (one formed by gravitational collapse) such outgoing radiation will always be present, having originated from the surface of the collapsing star.
Performing this calculation for the LBH using continuous streams of ingoing and outgoing radiation as done originally \cite{poisson} appears to be technically intractable. Fortunately we can make use of a model system that contains the same essential physics but is much easier to deal with technically. Specifically, we replace the continuous ingoing and outgoing streams by null thin shells that collide near the Cauchy horizon. Such an analysis was performed, for example, to show that mass inflation occurs in   rotating black holes \cite{collision} (though the rotating case was first considered using the Ori model \cite{chan}).  A further motivation for using this simplified model is that it allows us to make conclusions about the spacetime curvature without having to directly use the Einstein field equations. In fact the only reference to field equations that we will make is to assume that the Einstein equations hold to good accuracy far from the black hole; this is of course a well justified assumption. Not having to rely directly on field equations affords a significant advantage, because we wish  to test stability given a loop quantum gravity framework, a framework in which we have limited knowledge of the effective field equations that should be used. Of course this analysis is still semiclassical, but we should try to avoid using the classical field equations whenever possible, and the null shell model we will use here does just that.

\begin{figure}%[h]
 %\centering
  \includegraphics[width=5cm]{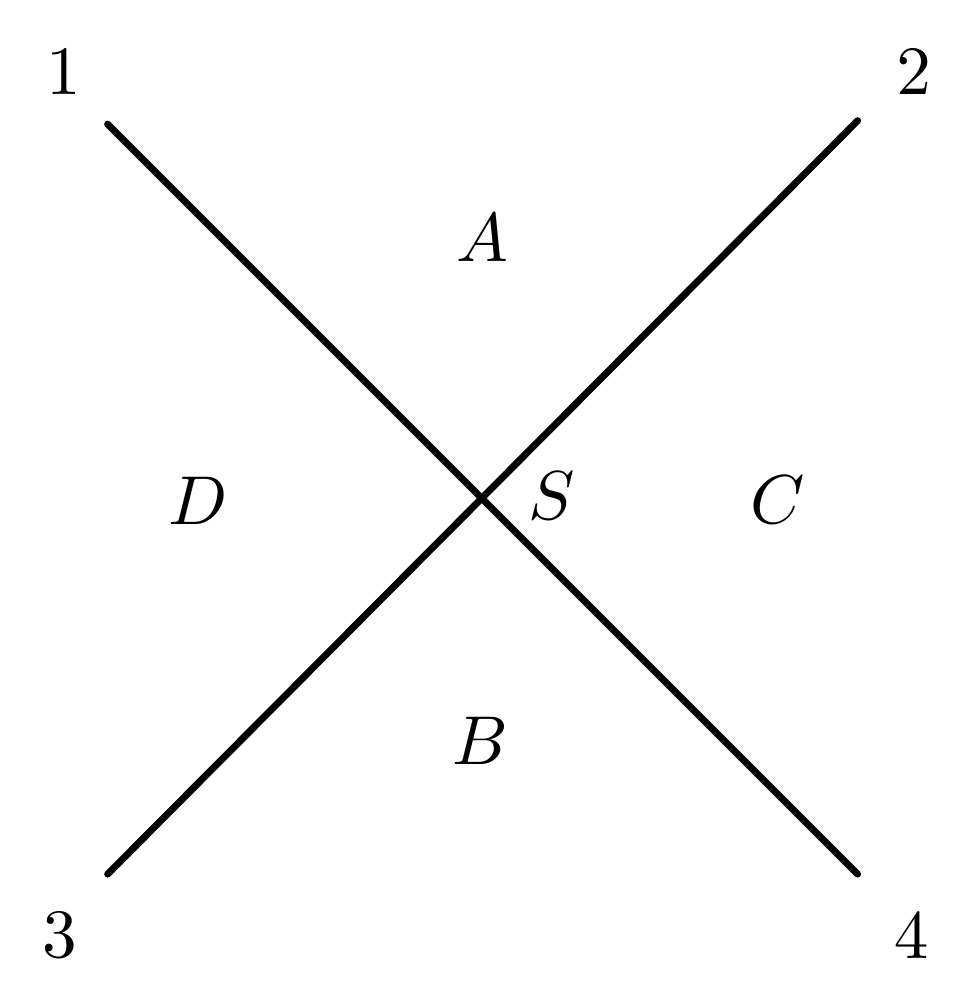}
    \caption{Two colliding null shells separate spacetime into four distinct regions: \(A\), \(B\), \(C\) and \(D\). In general the shells before the collision, 3 and 4, will not have the same properties as the shells after collision, 1 and 2. The two-sphere on which they collide is labeled \(S\). }
\label{DTRfig1}
  \end{figure}

Colliding null shells act to split the spacetime into four separate regions, as illustrated in Fig. \ref{DTRfig1} .
The DTR relation is a result that allows us to relate the metrics in each of these regions to each other (evaluated on the collision two-sphere \(S\)). The original DTR relation achieves this but is constrained to Schwarzschild spacetimes only \cite{origDTR}. The general DTR relation \cite{collision,thinshells}, whose derivation we briefly recapitulate,  evades this limitation.

While we  assume a spherically symmetric spacetime and shells that travel radially, the results presented are independent of this
latter assumption. For a given radial null shell, \(\Sigma\), we define the vector tangent to its generators
\begin{align}
    \ell^\alpha \equiv \frac{\partial x^\alpha}{\partial \lambda} , 
\end{align}
where \(\lambda\) parameterizes the shell; this parameter must be the same on both sides of \(\Sigma\) but is otherwise arbitrary and need not be affine on either side (in fact if the shell in question exerts transverse pressure then \(\lambda\) \emph{cannot} be made affine on both sides \cite{thinshells}). We also define the two tangent vectors \(e^\alpha_a\) that are transverse to the generators (latin indices run over two coordinates). Since \(\Sigma\) is null we have
\begin{align} \label{tangents}
    \ell_\alpha \ell^\alpha = 0 = \ell_\alpha e^\alpha_a. 
\end{align}
With our assumption of spherical symmetry the transverse vectors must lie in the two-sphere, and so can be chosen to be \(e^\alpha_\theta \partial_\alpha=\partial_\theta\) and \(e^\alpha_\phi \partial_\alpha=\partial_\phi\) without loss of generality. We can then define the effectively two-dimensional intrinsic metric on \(\Sigma\) by
\begin{align} \label{intmet}
    \sigma_{a b} \equiv g_{\alpha \beta}e^\alpha_a e^\beta_b .
\end{align}
The intrinsic metric for null shells is effectively two-dimensional because including the third tangent vector \(\ell^\alpha\) gives trivially zero entries: \(\sigma_{\lambda \lambda}=\sigma_{\lambda a}=0\) as can be seen from Eq. (\ref{tangents}); we therefore exclude these entries from the definition of the intrinsic metric.

We now define an object associated with \(\Sigma\),
\begin{align}
    K_{a b} \equiv \frac{1}{2}\mathscr{L}_\ell \left(\sigma_{a b}\right)=\frac{1}{2}\mathscr{L}_\ell \left(g_{\alpha \beta}\right)e^\alpha_a e^\beta_b ,
\end{align}
where \(\mathscr{L}_\ell\) is the Lie derivative along the vector \(\ell^\alpha\). We have dropped additional terms on the right-most side of the above equation since they evaluate to zero anyway, as a result of \(e^\alpha_a\) being transverse to \(\lambda\). We call the trace of this \(K\), and it is equal to the expansion rate of \(\Sigma\); we find
\begin{align} \label{trace}
    K \equiv \sigma^{a b}K_{a b}=\ell^\alpha_{\; ;\alpha}-\kappa , 
\end{align}
where \(\kappa\) is the ``acceleration" of the shell defined by \(\ell^\alpha_{\; ;\beta}\ell^\beta=\kappa \ell^\alpha\).

In order to consider what happens on the intersection two-sphere \(S\) we need to first make the principal assumption that the spacetime on \(S\) is no worse behaved than it is on the thin-shells individually. To understand what this means one must realize that by introducing an infinitely thin shell with nonzero stress-energy into our spacetime we are necessarily introducing a curvature singularity on the shell. This singularity simply arises as an artifact of the model and has a well understood physical interpretation. Despite this singularity, given a local coordinate system that is continuous across the shell the metric \(g_{\alpha \beta}\) remains continuous and piecewise continuously differentiable across the shell (the reader is referred to \cite{thinshells,text} for expositions on the thin-shell formalism). Thus, the assumption that the spacetime on \(S\) is no worse than it is on the shells individually is equivalent to requiring that for every point on \(S\) there is a local coordinate system in which the metric is continuous and piecewise continuously differentiable.

This assumption leads to a well defined notion of parallelism in vectors transverse to \(S\). In our model we have four null vectors that are orthogonal to \(S\): \(\ell_i^\alpha=\partial x^\alpha / \partial \lambda_i\), \(i=1,2,3,4\), where \(\lambda_i\) parameterizes shell \(\Sigma_i\). And since there are only two linearly independent null directions orthogonal to \(S\) the above assumption allows us to conclude that
\begin{align}
    \ell_1=a\ell_3 \, , \;\;\;\;\;\; \ell_2=b\ell_4 \, ,\nonumber
\end{align}
for some numbers \(a\) and \(b\). From this we immediately obtain
\begin{align} \label{parallel}
    (\ell_1\cdot \ell_2)(\ell_3\cdot \ell_4)=(\ell_1\cdot \ell_4)(\ell_2\cdot \ell_3) \;\;\;\; \text{on} \; S
\end{align}
This is a necessary geometric condition on the spacetime and is independent of any field equation.

Let us now define four functions on \(S\), \(Z_A\),...,\(Z_D\) as, for example
\begin{align} \label{dtr1}
    Z_A \equiv \frac{K_1 K_2}{\ell_1 \cdot \ell_2}, 
\end{align}
where \(K_i\) is the expansion of shell \(\Sigma_i\), as defined in Eq. (\ref{trace}). The other functions are similarly defined, where the subscript on \(Z\) corresponds to the spacetime region bounded by the two shells referred to on the right-hand-side. Of principle importance is the fact that these functions are independent of the parameters \(\lambda_i\) used to describe the shells. This is easily seen because under reparameterization \(\lambda_i \rightarrow \tilde{\lambda}_i\) the expansion \(K_i\) rescales the same as \(\ell_i\). This allows us to put Eq. (\ref{parallel}) into a parameter-independent form
\begin{align} \label{dtr2}
    |Z_A Z_B|=|Z_C Z_D| \;\;\;\; \text{on} \; S \,  ,
\end{align}
where the absolute value bars are necessary because in using the \(Z\) functions we have lost sign information.
For example, \(Z_A\) is invariant under the reparameterization \(\lambda_1 \rightarrow -\lambda_1\), unlike \(\ell_1\).

This is what is known as the generalized DTR relation. It allows us to relate the metric in the four spacetime quadrants (on \(S\)) in a way that is independent of the parameters \(\lambda_i\) used to generate the shells and indeed is completely independent of the stress-energy properties of the shells. That is, nowhere in the derivation of this result have we used Einstein's field equations; rather the relation is a necessary geometric condition of the spacetime.

\section{Mass Inflation}
\label{massinflation}

Classically, mass inflation is a process in which the mass parameter of the black hole diverges unboundedly at the Cauchy horizon.  The outgoing radiation acts to separate the inner apparent horizon of the black hole from its Cauchy horizon, and this separation is  a necessary condition for mass inflation to occur. The effect is induced independent of whatever form the outgoing radiation may take; the only requirement is that it has a nonzero energy (necessarily negative energy since it is outgoing). That is, we expect the radiation to change the mass parameter of the black hole, thus creating the separation between the apparent and Cauchy horizons stated above. This process was originally explicated for  Reissner-Nordstr\(\ddot{\text{o}}\)m black holes  \cite{poisson} and then simplified in a model constructed by  Ori \cite{ori} (we will discuss this model of mass inflation in the Appendix). From a more physical viewpoint, the inability of the counter streams of radiation to locally travel faster than the speed of light relative to each other can be seen to produce arbitrarily large amounts of gravitational energy as the Cauchy horizon is approached, and this energy manifests itself in the material form of arbitrarily large local mass. Hamilton and Avelino provide an excellent discussion of the precise physics at work here \cite{hamilton} and they note that the simplified models of mass inflation typically used (including the one we employ) do well at accurately capturing the results obtained in their more comprehensive study.

\begin{figure}%[h]
% \centering
  \includegraphics[width=7.0cm]{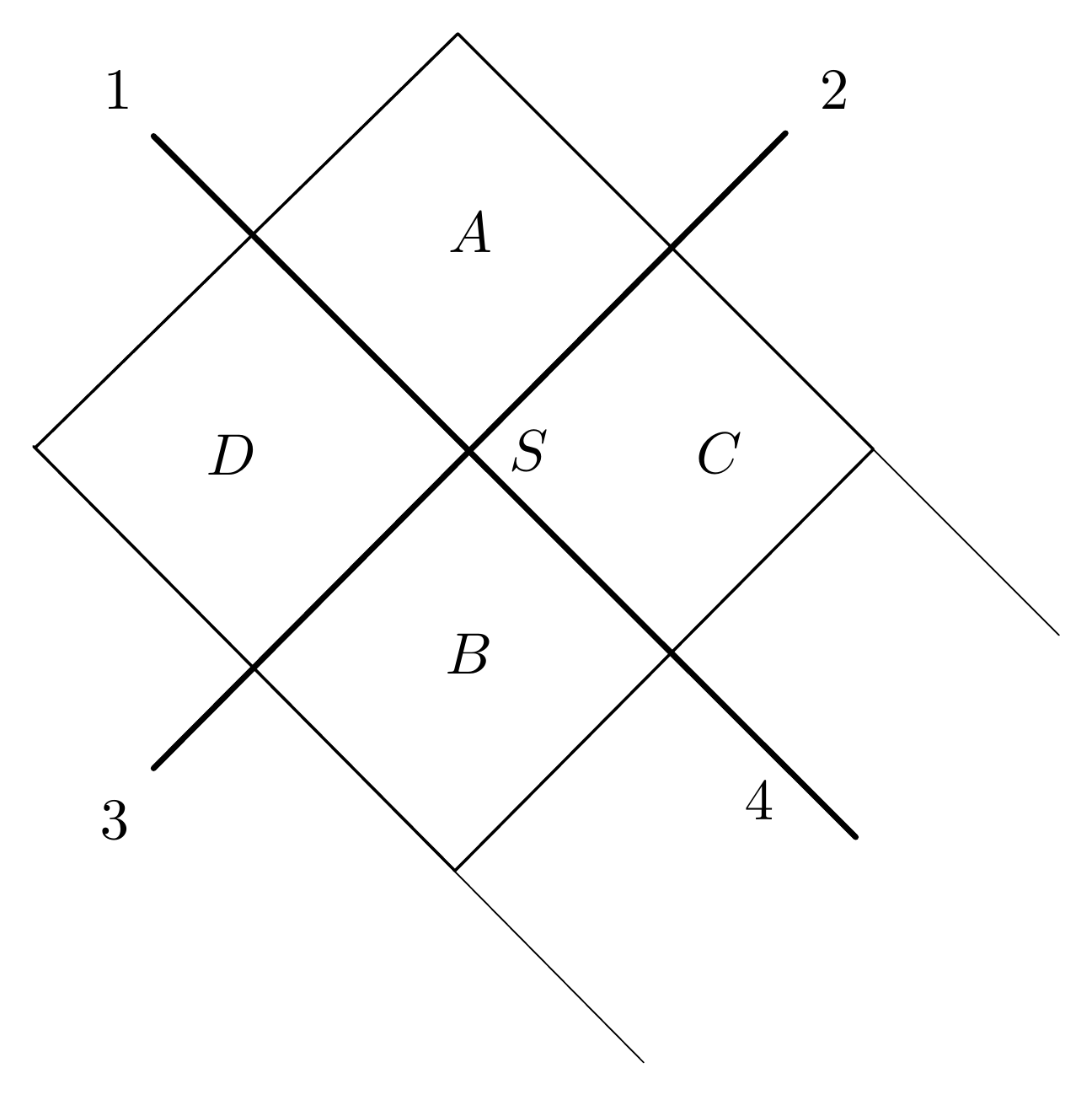}
    \caption{The colliding null shells split the loop spacetime into four regions, the metrics in each of which can be related via the generalized DTR relation. We consider the scenario in which the ingoing shell is arbitrarily close to the Cauchy horizon. }
\label{DTRfig2}
  \end{figure}

The calculation that we will perform models the ingoing and outgoing radiation as null thin shells that collide near the Cauchy horizon, as displayed in Fig. \ref{DTRfig2}. This model captures the same essential physics present in the continuous scenario while making the problem much simpler to solve. Furthermore, it allows us to make conclusions without the need to rely on specific field equations, the form of which are not known given a loop quantum gravitational framework.
 The colliding shells act to split the spacetime into four distinct regions, each equipped with their own metric. A principal assumption we make is that each of the spacetime quadrants is described by a loop metric of the form given by Eq. (\ref{g}), each  with a different mass parameter \(m\). This assumption seems reasonable considering that the loop metric was derived as a modified spherically symmetric vacuum solution, and each of the four regions are themselves representative of a spherically symmetric LBH vacuum spacetime. It is also supported by the requirement that outside the black hole classical general relativity holds to a good approximation; in a Schwarzschild black hole endowed with an ingoing thin-shell we know that the mass parameter is constant within the past and future regions of the shell. Thus \(m\) in the loop black hole can also be expected to have this behaviour for large \(r\). Physically we expect  stationarity to hold, meaning that \(m\) should be independent of the advanced time \(v\) within this region. Hence in each quadrant there will be only small deviations from constant \(m\).

With this assumption we can use the generalized DTR relation, Eqs. (\ref{dtr1}) and (\ref{dtr2}), to say how the mass parameters in each region, \(m_A\), \(m_B\) etc, are related to each other on the collision two-sphere \(S\). Since the DTR relation itself only gives us information on \(S\),  this assumption allows us to say that the parameter \(m_B\), for example, takes the same value on \(S\) as it does in the rest of the \(B\) region up to negligibly small corrections.

We wish to find how the relation   (\ref{dtr2}) evaluates given that each region is described by the loop metric. Working in \((t,r,\theta,\phi)\) coordinates the tangent vectors \(e^\alpha_a\) transverse to the generators of the radially traveling shells are given by \(e^\alpha_\theta=(0,0,1,0)\) and \(e^\alpha_\phi=(0,0,0,1)\). The intrinsic metric on the shells, Eq. (\ref{intmet}), is therefore given by
\begin{align}
    \sigma_{a b}=H(r)
    \begin{bmatrix}
    1 & 0 \\
    0 & \sin^2\theta
    \end{bmatrix}
    , \;\;\;\;\; H(r)=r^2+\frac{a_0^2}{r^2} , 
\end{align}
where \(a_0\) is interpreted as the minimum area of the theory. Recall that \(\ell^\alpha\) is the vector tangent to the generators of the shell in question. The entries of the intrinsic metric are scalars with respect to the full four dimensional spacetime, and so when we compute the Lie derivative of them along the vector \(\ell^\alpha\) we simply use a partial derivative:
\begin{align}
    K_{a b} \equiv \frac{1}{2}\mathscr{L}_\ell \left(\sigma_{a b}\right)=\frac{1}{2} \ell^\alpha \left(\sigma_{a b}\right)_{,\alpha} . 
\end{align}
From this it is easily seen that the trace, Eq. (\ref{trace}), takes the form
\begin{align}
    K=\frac{1}{H}\ell^\alpha H_{,\alpha} , \nonumber
\end{align}
or rather, when referring to one of the specific shells \(\Sigma_i\) we have
\begin{align}
    K_i=\frac{1}{H}\ell^\alpha_i H_{,\alpha}. 
\end{align}
We now obtain the form of the quantities \(Z_A\), \(Z_B\) etc, as defined by Eq. (\ref{dtr1}). For example we have
\begin{align} \label{za}
    Z_A=\frac{1}{H^2}\frac{\ell^\alpha_1\ell^\beta_2H_{,\alpha}H_{,\beta}}{\ell_1\cdot \ell_2} , 
\end{align}
where \(\ell^\alpha_1\) and \(\ell^\alpha_2\) are the vectors tangent to the generators of shells \(\Sigma_1\) and \(\Sigma_2\) that bound spacetime region \(A\).

To relate \(Z_A\) to the metric \(g^{\alpha \beta}_{(A)}\) in region \(A\) we make use of the completeness relation
\begin{align} \label{completeness}
    g^{\alpha \beta}_{(A)}=\frac{2\ell^{(\alpha}_1\ell^{\beta)}_2}{\ell_1\cdot \ell_2}
    +\sigma^{a b}e^\alpha_a e^\beta_b . 
\end{align}
This can be verified by computing all inner products between \(\ell^\alpha_1\), \(\ell^\alpha_2\) and \(e^\alpha_a\). If we then contract the metric with \(H_{,\alpha} H_{,\beta}\) the result is seen to be
\begin{align}
    g^{\alpha \beta}_{(A)}H_{,\alpha}H_{,\beta}=\frac{2 H^2 K_1 K_2}{\ell_1\cdot \ell_2} , 
\end{align}
where we have used \(e^\alpha_a H_{,\alpha}=0\), since that \(H_{,\alpha}=0\) for all \(\alpha \neq r\). Comparing this result with Eq.(\ref{za}) we immediately obtain
\begin{align}
    Z_A=\frac{1}{2H^2}g^{\alpha \beta}_{(A)}H_{,\alpha}H_{,\beta}=\frac{1}{2H^2}g^{r r}_{(A)}(H_{,r})^2 .
\end{align}
For the loop black hole we have \(H=r^2+a_0^2/r^2\), and so
\begin{align} \label{za2}
    Z_A=\frac{2}{r^2}\frac{(r^4-a_0^2)^2}{(r^4+a_0^2)^2}F_A . 
\end{align}
where \(F_A\) is the metric function \(g^{rr}\) in region \(A\). Under the assumption that the region is described by the loop metric this function is given by
\begin{align} \label{Fa}
    F_A=\frac{(r-2m_A)(r-2m_AP^2)r^4}{(r+2m_AP)^2(r^4+a_0^2)} , 
\end{align}
where \(m_A\) is the mass parameter in region \(A\). The forms of \(Z\) in the other regions follow analogously to this.

Note that we have assumed continuous \(r\) at \(S\); this results from our assumption in Sect. \ref{DTR} that the spacetime at \(S\) is no worse behaved than it is across the shells. We know that \(r\) must be continuous across the shells in order for them to have a well defined intrinsic geometry (specifically, \(\sigma_{a b}\) must be continuous across the shells) and so the assumption that \(r\) is continuous on \(S\) follows directly from the assumption already made in deriving the generalized DTR relation.

Recall that the generalized DTR relation tells us that the four regions of spacetime, as seen in Fig. \ref{DTRfig2}), are related at \(S\) by \(|Z_A Z_B|=|Z_C Z_D|\). From Eq. (\ref{za2}), the form of the metric function \(F\) and the assumption of continuous \(r\) we finally obtain the defining equation of this analysis:
\begin{align} \label{maineq}
    |X_A X_B|=|X_C X_D|, 
\end{align}
where for convenience we have defined
\begin{align}
    X_A \equiv \frac{(r-2m_A)(r-2m_AP^2)}{(r+2m_AP)^2}
\end{align}
and similarly for \(X_B\) etc. This equation gives us a necessary condition on the four mass parameters

We now wish to ask the question of what happens when the ingoing shell is arbitrarily close to the Cauchy horizon (equivalently, the shell enters the black hole arbitrarily far in  the future as seen by an outside observer). In order to do this we must specify how the energy of the shell should depend on its distance to the Cauchy horizon or, more precisely, how it should depend on the advanced time \(v\). In Sect. \ref{preliminary} we discussed how, in the case of continuous radiation, the energy density of the ingoing radiation should fall off as an inverse power, \(\sim v^{-\gamma}\) where \(\gamma \geq 12\), in the large \(v\) limit (recall that the Cauchy horizon is located at \(v=\infty\)). This occurs as a result of backscattered gravitational radiation, and is considered a very generic form of matter present in realistic black hole spacetimes. In classical spherically symmetric black holes this ingoing radiation acts to increase the mass of the black hole as a function of \(v\) such that the change in mass is proportional to the power law decrease expressed by the energy density; i.e. \(m'(v) \propto v^{-\gamma}\). This  classical result relies on Einstein's field equations, and so cannot necessarily be expected to hold in the loop black hole. However, it is known that far from the loop black hole the spacetime is negligibly deviant from that of a classical Schwarzschild black hole and that Einstein's equations are expected to hold to a good approximation in this region. Therefore, under the assumption that the mass parameter \(m\) of the loop metric is fixed along a line of constant \(v\) (in the null shell model this corresponds to assuming fixed \(m\) within a given quadrant) it means that the classical result of \(m'(v) \propto v^{-\gamma}\) will hold to a good approximation in the loop black hole as well. When it comes to the null shell model, however, we must still make a choice about how to model the energy of the ingoing shell as a function of \(v\). This leads us to our  third primary assumption:  that the energy of the ingoing shell, and therefore the jump in mass parameters across the shell, follows this same power law
\begin{align} \label{massjump}
    m_C-m_B \propto v^{-\gamma} .
\end{align}
This is the most natural condition that can be placed on the shell energy given the nature of the model being considered, and it is what will most accurately capture the important physics of the full system. The reader should note that this is the one and only place that any reference to the Einstein field equations is made, and it is simply the statement that the field equations should hold when far from the black hole, which indeed they should.

There is a source of possible confusion that should be addressed.  Previously we  stated that the mass parameters should be fixed within their given regions, and yet Eq. (\ref{massjump}) seems to imply that they are changing. What must be understood is that the action of ``moving the ingoing shell closer to the Cauchy horizon" really just means that we are thinking of a series of distinct spacetimes. A given spacetime will have the ingoing shell located at some fixed value of \(v\), and in our model the  difference in
mass parameters \(m_C-m_B\) depends on this value via Eq. (\ref{massjump}). With this framework we are going to consider what happens when the spacetime in question has its shell located at an arbitrarily large value of \(v\), but the mass parameters of this spacetime are still constant within their respective quadrants. For the rest of this section when something is said to be a function of \(v\) it is implied that \(v\) is the value of advanced time at which the shell is located, and by increasing \(v\) we moving through a sequence of spacetimes. The only way in which the masses can be said to change as a function of \(v\) is within this context.

With this understood, it must be specified that \(m_B\), the mass of the black hole as detected by an outside observer prior to the shell entering the hole, is to be considered ``fixed" as we cycle through different spacetimes (as the shell's value of \(v\) is increased) and the mass after the shell has entered, \(m_C\), will ``change" over the series of spacetimes. Specifically, as a function of \(v\) the future mass is seen to decrease as \(m_C=m_B+v^{-\gamma}\). This choice is both the most natural (as opposed, for example, to considering \(m_C\) as fixed and \(m_B\) as increasing) and it allows us to use a well defined choice of advanced time \(v\). That is, the advanced time we use will be that which is defined with respect to the \(B\) region of the spacetime. Equivalently, the choice of fixed \(m_B\) tells us that that the Cauchy horizon will be located exactly at \(r=r_-\) and \(v=\infty\), where
\begin{align}
    r_-=2m_BP^2 . 
\end{align}
In addition, the metric functions of the \(B\) region decay exponentially as the horizon is approached, as demonstrated in Sect. \ref{preliminary}. That is,
\begin{align}\label{XBexp}
    X_B \propto -e^{-\kappa_- v}, 
\end{align}
where \(\kappa_-\) is the surface gravity of the Cauchy horizon; a positive quantity that is defined in Eq. (\ref{surfgrav}) (with \(m\) replaced by \(m_B\)).   The negative sign in front of the exponent  indicates that we are approaching the Cauchy horizon from between the horizons, and so \(X_B\) will be negative as this approach takes place. This fact is not as obvious as it may seem for the loop black hole. The only reason it is correct is because \(m_B\) is fixed; this subtlety will be discussed later in this section and in the Appendix.

In the same way that \(m_C-m_B\) will be the energy of the ingoing shell, the mass difference \(m_D-m_B\) will be equal to (the negative of) the energy of the outgoing shell, at least in a classical setting. That is, the outgoing shell has negative energy density and we have \(m_D>m_B\). However, this condition results from the Einstein field equations, and in the case of the outgoing shell we cannot necessarily make the same conclusion because we don't have the extension to the outside of the black hole that we do for the ingoing shell (i.e. the emission of the outgoing shell from the collapsing star occurs inside the black hole). This means that without knowledge of the effective field equation we can not know with certainty whether \(m_D\) will be greater than or less than \(m_B\). However the loop black hole admits a maximally extended universe, suggesting that \(m_D>m_B\) in the loop black hole, as in the classical case.

Fortunately, whatever the case, all we need to know is that \(m_D \neq m_B\), because this is what produces the separation between the apparent and Cauchy horizons that is necessary for mass inflation. We will soon see explicitly why we don't need to specify the direction of mass change across the shell.

From Eq. (\ref{maineq}) we have
\begin{align} \label{maineq2}
    |X_A|=\left|\frac{X_C}{X_B}\right||X_D| . 
\end{align}
Using this we are going to examine what happens to \(X_A\) when we take \(v\rightarrow \infty\), and we will find in fact that it grows unboundedly large. This is not obvious because while \(X_B\) approaches zero as \(v \rightarrow \infty\), so does \(X_C\) (after all we have \(X_B=X_C\) at \(v=\infty\)). The point is that \(X_B\) decays faster than \(X_C\) as the Cauchy horizon is approached, and so the limit of \(X_C/X_B\) will diverge. More explicitly, given \(m_C=m_B+v^{-\gamma}\) we have that
\begin{align} \label{dtrresult}
    X_C &\equiv \frac{(r-2m_C)(r-2m_CP^2)}{(r+2m_CP)^2} \nonumber \\
    &=\frac{(r-2m_B)(r-2m_BP^2)-2y(r)v^{-\gamma}}{(r+2m_BP)(r+2m_BP+4Pv^{-\gamma})}+\mathcal{O}(v^{-2\gamma}) \nonumber \\
    &\simeq \frac{(r+2m_BP)^2X_B-2y(r)v^{-\gamma}}{(r+2m_BP)^2} , 
\end{align}
for large \(v\). Here we have used
\begin{align}
    y(r) \equiv (1+P^2)r-4m_BP^2 . 
\end{align}
Thus, as \(v \rightarrow \infty\) we see
\begin{align}
    \frac{X_C}{X_B} &\sim -y(r_-)v^{-\gamma}X_B^{-1} \\
    &\sim y(r_-)v^{-\gamma}e^{\kappa_- v} , 
\end{align}
which indeed diverges.

The assumption that the outgoing shell produces a mass difference \(m_D \neq m_B\) means that \(X_D\) is nonzero on the Cauchy horizon, and so by Eq. (\ref{maineq2}) we conclude that \(X_A\) diverges as
\begin{align}
    |X_A| \sim v^{-\gamma}e^{\kappa_- v}, \;\;\;\; v \rightarrow \infty . 
\end{align}
As stated above, this is independent of whether \(m_D\) is greater than or less than \(m_B\) since this only acts to change the sign of \(X_D\) on the Cauchy horizon. Thus, we see that indeed there is divergence at the Cauchy horizon. In addition, this is the exact same form of divergence that is observed in classical mass inflation.  Note that this divergence is also similar to the form that was observed from the perturbation calculation in Sect. (\ref{preliminary}).

As a quick aside, we find that \(y(r_-)=y(2m_BP^2)=4m_BP^2(P^2-1)<0\) since \(P<1\). This tells us that \(X_C/X_B<0\) for large \(v\) which, since \(X_B\) remains negative, tells us that \(X_C\) is \emph{positive} as the Cauchy horizon is approached. Recall our earlier remark noting that if the loop black hole is subjected to ingoing radiation the metric function need not remain negative for large \(v\)  before the Cauchy horizon is reached. This is a result of the fact that, unlike the Reissner-Nordstr\(\ddot{\text{o}}\)m black hole, the inner horizon of the loop black hole \emph{inflates} with increasing mass. This produces a strange effect where, when continuous ingoing radiation is present, an outgoing geodesic can pass by the radius at which the Cauchy horizon \emph{will be} located. The geodesic will then reach the apparent horizon, turn around, and straddle the apparent horizon before settling on to Cauchy horizon at \(v=\infty\).  We discuss this situation more thoroughly in the Appendix. The positivity of \(X_C\) observed here is the shell model analog of this phenomenon. This has no consequence on the divergent behaviour of  \(X_A\).

We have seen that \(X_A\), and therefore the metric function \(F_A\) diverges at the Cauchy horizon.  Similar reasoning yields the same conclusion for  the Reissner-Nordstr\(\ddot{\text{o}}\)m black hole, except that it is the function \(f_A=1-2m_A/r+e^2/r^2\) that diverges, thus indicating that the mass \(m_A\) diverges, hence the name mass inflation.  In the loop black hole, however,   \(F_A \rightarrow \infty\) cannot be achieved from a diverging mass parameter  (see Eq. (\ref{Fa})). Indeed, \(F_A\) remains perfectly finite for \(m_A \rightarrow \infty\), and so  referring to  our result as mass inflation is arguably a little disingenuous. However for historical reasons and because our result follows from the same physics and has the same characteristics as classical mass inflation, we will continue to call it as such.

It is not hard to see that the only way for \(F_A\) to actually diverge  is for the mass to take a negative value of \(m_A \rightarrow -r_-/2P\). It is not clear how to interpret this  strange result.  We interpret the parameter \(m\) in the loop metric as being the mass because at infinity \(M_{ADM}=(1+P)^2m\), where $M_{ADM}$ is the classical ADM mass. To the future of the outgoing shell, however, the parameter \(m\) no longer has this classical  interpretation. Alternatively, one could  argue that we should define $m$ in the context of a quasi-local mass computed
from $F$.  However this interpretation is based on the Einstein field equations \cite{quasimass}, and thus the same interpretation can certainly not be applied inside the loop black hole. We therefore simply consider \(m\) to be a parameter of the spacetime, and so are not concerned with the possibility of negative \(m_A\), since  this negativity occurs in a region of spacetime where there is no good reason to interpret \(m_A\) as being a mass anyway; the result is simply a geometric requirement.

When the limit \(m_A \rightarrow -r_-/2P\) is taken at the Cauchy horizon the Ricci and Kretschmann scalars computed from the loop metric go as \(R \sim (r_-+2m_AP)^{-4}\) and \(K \sim (r_-+2m_AP)^{-6}\) respectively, thus indicating that they diverge in this limit. That is, the phenomenon seen to occur here causes a scalar curvature singularity at the Cauchy horizon, the same result as that obtained in classical mass inflation. However the DTR
analysis we have performed \emph{only} tells us what happens to \(F_A\). Since for the loop metric \(F_A\) is a quadratic in \(m_A\) there are two possible solutions for \(m_A\), yielding an ambiguity in how \(m_A\) approaches this negative value as, say, a function of \(v\). In the Reissner-Nordstr\(\ddot{\text{o}}\)m black hole there is no ambiguity since the metric function is linear in the mass. In our case all we know is that \(F_A\) increases, eventually diverging at the Cauchy horizon. Solving for \(m_A(F_A)\) we find that as \(F_A\) increases one of the solutions goes negative and approaches \(-r_-/2P\) continuously.  In the other solution  \(m_A\) first goes to positive infinity and then jumps   to negative infinity before approaching  \(-r_-/2P\) from the negative side.

\begin{figure} %[h]
%  \centering
 \subfloat{\label{Fvm1}
  \includegraphics[width=0.4\textwidth]{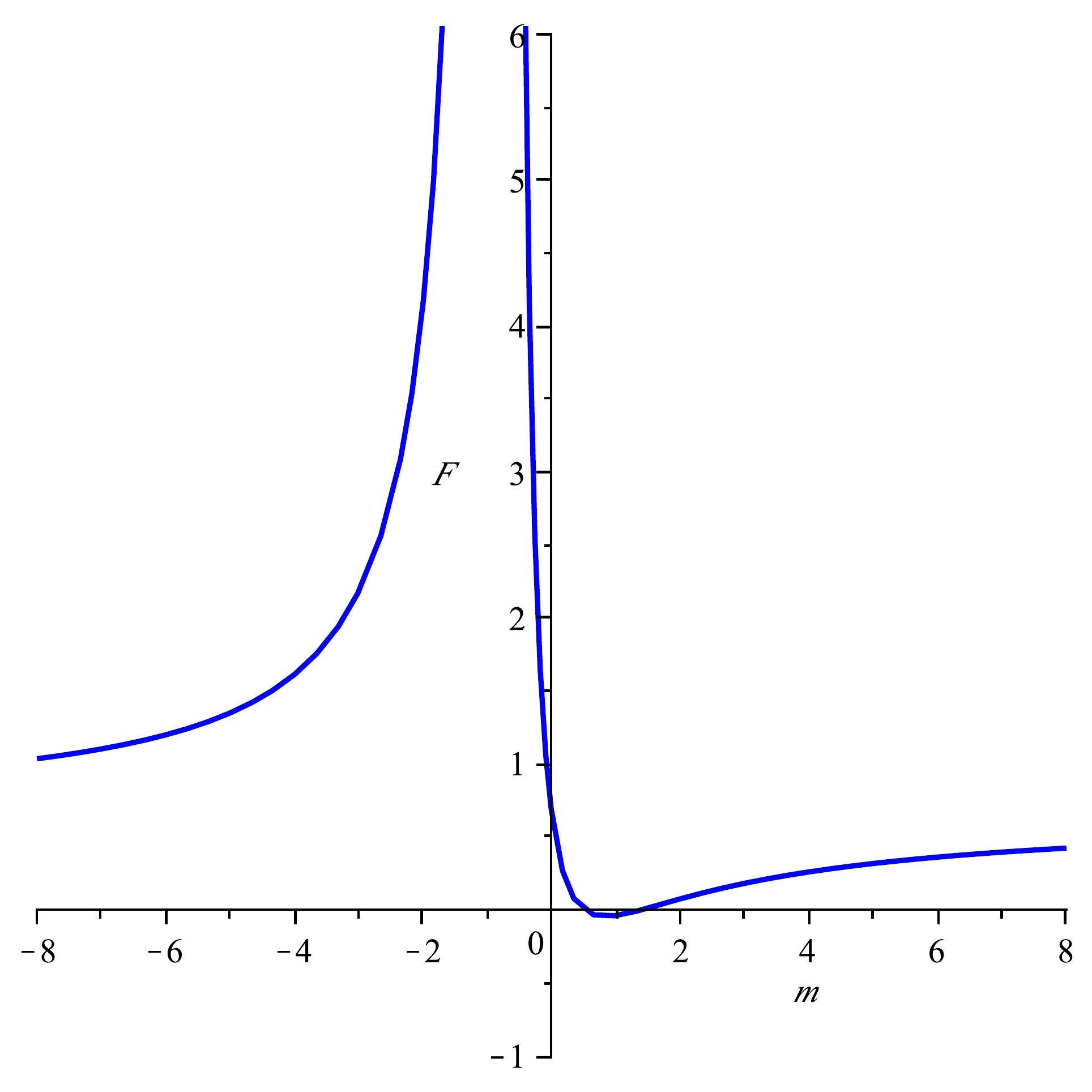} }
  \;
  \subfloat{\label{Fvm2}
  \includegraphics[width=0.4\textwidth]{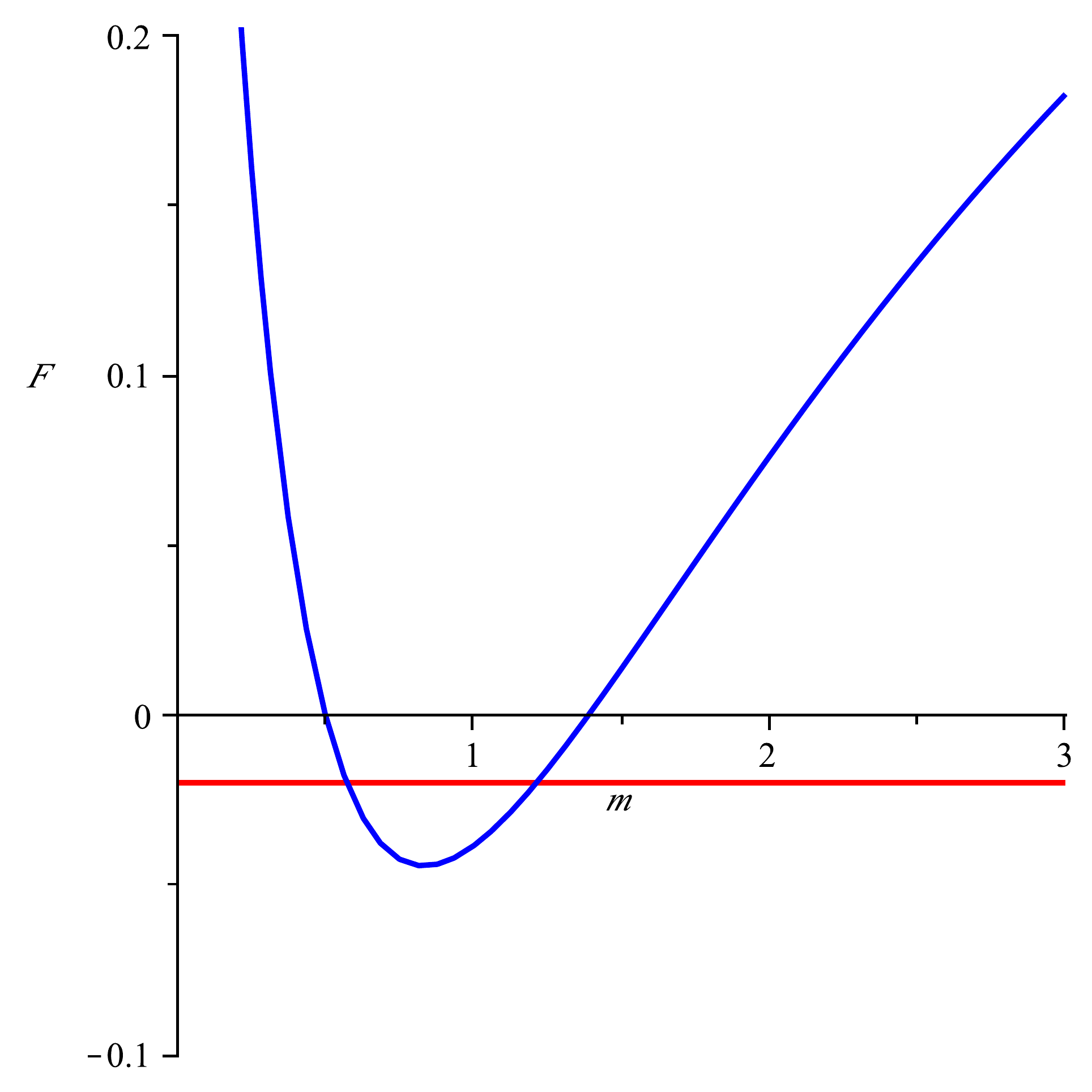} }
  \caption{\(F_A\) plotted as a function of \(m_A\) with fixed values \(r=1\), \(P=0.6\) and \(a_0=0.5\). The top figure gives a large view of the function; note that \(F_A\) diverges at \(m_A=-r/2P=-5/6\). Although it isn't obvious from the figure, \(F_A\) asymptotes to a value of \(r^4/(r^4+a_0^2)=0.8\) when \(m_A \rightarrow \pm \infty\). The bottom figure displays the range of \(m_A\) from 0 to 3 and we include a horizontal line to emphasis that, given a value of \(F_A\), there are two possible values of \(m_A\). As \(F_A\) increases we see that \(m_A\) will either decrease or increase depending on whether the initial value is taken to be on the left or right side of the minimum in \(F_A\), respectively.}
  \label{Fvm}
\end{figure}

To understand how these two solutions arise we plot at fixed $r$ in Fig.\ref{Fvm} \(F_A\) as a function of \(m_A\), where \(P\) and \(a_0\) are some fixed values.  We see that, given some initial value of \(F_A\) (presumably negative), there are two possible values of \(m_A\); either on the left or the right side of the minimum in \(F_A\). The DTR relation tells us that \(F_A\) increases as the Cauchy horizon is approached, and we see from the figure that as \(F_A\) is increased the value of \(m_A\) will either decrease or increase depending on which option we choose for its initial value. If it is to the left of the minimum in \(F_A\) its value will decrease continuously and eventually reach \(-r/2P\) as \(F_A\) diverges. If it is to the right of the minimum, however, its value will increase and diverge to infinity when \(F_A=r^4/(r^4+a_0^2)\) before reemerging at negative infinity and approaching \(-r/2P\) from the negative direction.

Note that, of course, \(F_A\) is also a function of \(r\) and so the plot will change as the evolution occurs since the value of \(r\) will also be changing. We find that the same qualitative behaviour occurs in the full scenario as well. That is, along an outgoing null geodesic \(r\) goes like \(r(v)=r_-+\kappa_-^{-1}e^{-\kappa_- v}\) for large \(v\) (see Sect. \ref{preliminary}), and this can be substituted into Eq. (\ref{dtrresult}). When this is done and the resulting quadratic is solved, we find that  the two solutions  follow the behaviour described above.

Thus, we see that which evolution of \(m_A\) actually takes place depends critically on the initial condition placed on \(m_A\). Unfortunately within the confines of our model we have no way of specifying what this initial condition should be. This is because the power law decay given by Eq. (\ref{massjump}) is only valid in the large \(v\) limit. Presumably there will be a nonzero influx of radiation during early times which we have no knowledge of, and so \(m_A\) will have already gone through some evolution by the time the large \(v\) regime is reached. Despite this limitation, however, we can perform a simple thought experiment in which we suppose that there is some point in the past, shortly after the collapse of the star, at which the influx starts. That is, we imagine that for some \(v=v_0\) the energy density of the ingoing radiation transitions continuously from zero to nonzero. In the thin-shell picture, this initial ingoing radiation will intersect the outgoing radiation shell at some \(r=r_0\), producing an intersection two-sphere located at \(S_0=(v_0,r_0)\). In this scenario we have, for example, that \(m_C(v=v_0)=m_B\) because the black hole has only just begun to accrete extra mass. In addition, it is clear from physical considerations and the assumption that \(m_A\) should evolve continuously that we also have \(m_A(v=v_0)=m_D\), thus giving us an initial condition for \(m_A\). From here, in order to determine which evolution of \(m_A\) will ensue we note that the minimum in \(F_A\) seen in Fig. \ref{Fvm}) is located at mass \(m_A=r/2P\). Therefore, we see that if \(m_A(v_0)=m_D>r_0/2P\) then \(m_A\) will follow the evolution that takes it to infinity and back around. Alternatively, if \(m_D<r_0/2P\) then \(m_A\) will proceed to its final value of \(-r_-/2P\) continuously. Given the scenario in our thought experiment, the evolution of \(m_A\) is thus entirely dependent on the value of \(m_B\), whether \(m_D\) is greater than or less than \(m_B\) and by how much, and the radius \(r_0\) at which the cross streams first intersect.

Note that   for \(m \rightarrow \pm \infty\) the metric functions approach \(F \rightarrow r^4/(r^4+a_0^2)\) and \(G \rightarrow 16m^4P^4/(r^4+a_0^2) \rightarrow +\infty\), yielding a diverging volume element \(\sqrt{-g}\). However  both the Ricci and Kretschmann scalars, \(R\) and \(K\), remain perfectly finite for diverging \(m\), a fact that remans true when we include a functional dependence \(m=m(v)\); i.e. the diverging derivatives of \(m(v)\) do not spoil the fact that \(R\) and \(K\) are finite.

Furthermore we find that there are no nonscalar singularities in this limit.
The presence of a nonscalar singularity is signalled by the divergence of at least one of the tetrad components of the Riemann curvature tensor when the tensor is evaluated in a parallely propagated orthonormal frame (PPON frame) \cite{sing}. If all of the components remain finite this means that the observer traveling along the geodesic defining the PPON frame will not experience any diverging tidal or inertial forces upon approaching the point in question. Given a radial geodesic \(U^\alpha\), we construct a PPON tetrad by letting the first vector be tangent to this geodesic \(e_1^\alpha=U^\alpha\) and, given spherical symmetry, we allow two of the other vectors \(e_3\) and \(e_4\) to lie in the two-sphere, at which point \(e_2\) follows directly from orthonormality \(\eta_{i j}=e_i^\alpha e_j^\beta g_{\alpha \beta}\). A radial geodesic (inside the black hole) on the loop metric takes the form
\begin{align}
    \hspace{-0.2cm} 
    U^t=\frac{E}{G}, \;\;\;\; U^r=-\sqrt{\frac{F}{G}(E^2-G)}, \;\;\;\; U^\theta=U^\phi=0 , 
\end{align}
where \(E\) is a constant (see Sect. \ref{preliminary}). Thus, in \((t,r,\theta,\phi)\) coordinates one finds the PPON frame generated by this geodesic to be
\begin{align} \label{PPON}
    e_1^\alpha&=\left(\frac{E}{G},\; -\sqrt{\frac{F}{G}(E^2-G)}, 0, 0\right), \nonumber \\
    e_2^\alpha&=\left(\frac{1}{G}\sqrt{E^2-G},\; -E\sqrt{\frac{F}{G}},0,0\right), \\
    e_3^\alpha&=\left(0,0,\frac{1}{\sqrt{H}},0\right), \;\;\;\; e_4^\alpha=\left(0,0,0,\frac{1}{\sqrt{H}\sin\theta}\right) .\nonumber
\end{align}
Projecting the Riemann curvature tensor on to this basis, \(R_{i j k \ell}=e_i^\alpha e_j^\beta e_k^\gamma e_\ell^\delta R_{\alpha \beta \gamma \delta}\) we find, somewhat miraculously, that \emph{all} components remain finite when \(m\) is taken to infinity.

These results suggest that taking the limits \(m\rightarrow \pm \infty\) does not result in a curvature singularity. Note that, like the scalar invariants, the curvature appears to remain finite even when we include a functional dependence \(m=m(v)\) (note that the tetrad was evaluated with respect to \((v,r,\theta,\phi)\) coordinates when performing this calculation). Of course in this dynamical case the tetrad in Eq. (\ref{PPON}) is no longer a PPON frame, and unfortunately finding a closed form for radial geodesics in this spacetime is nontrivial since \(t\) is no longer a Killing direction. Even so, when \(m=m(v)\) then Eq. (\ref{PPON}) still represents the frame of an accelerated observer, and the fact that the tetrad components \(R_{i j k \ell}\) remain finite when \(m(v)\) and its derivatives diverge means that, at least for this accelerated observer, the spacetime curvature appears to remain finite. Note that \(R_{i j k \ell}\) also remain finite if we use, for example, the static orthonormal tetrad given by \(e_1^\alpha=(1/\sqrt{G},0,0,0)\), \(e_2^\alpha=(0,\sqrt{F},0,0)\), along with \(e_3^\alpha\) and \(e_4^\alpha\) given as they are in Eq. (\ref{PPON}).

To summarize,  we have been unable to find a way in which the loop metric becomes singular when \(m\rightarrow \pm \infty\), apart from a divergence in the volume element, a coordinate artifact. This, coupled with the fact that the metric functions \(F\) and \(G\) take the same value when \(m\) is both positive and negative infinity (since \(G \propto m^4\) for large \(\pm m\) , then \(G\) attains the same limit  for either sign), suggests that evolution of
 \(m_A\) through infinity to \(-r_-/2P\)  may indeed represent a perfectly well behaved and continuous evolution of spacetime. Note that \(m_A\)  is a parameter of the spacetime that is only accessible deep inside the black hole and is not observable from the outside. Furthermore, we can't identify \(m_A\) with any sort of quasi-local mass since the Einstein field equations are not expected to hold in this region of spacetime. We thus identify \(m_A\) only as being a parameter of the internal spacetime, one
whose evolution yields a   continuous sequence of spacetimes  even though the observed evolution of \(m_A\) itself is discontinuous.

Whichever solution \(m_A\) follows, it will always limit to the value of \(-r_-/2P\) as the Cauchy horizon is approached, and this certainly \emph{is} a surface of infinite curvature. It is interesting to note, given that the spacetime appears regular for \(m \rightarrow \pm \infty\), that the DTR analysis ferrets out  the only way that the loop metric \emph{knows how} to become singular.

The model used and the DTR analysis performed here are quite general. Any black hole with a Cauchy horizon -- classical or quantum -- will be hard pressed to avoid the same result.

The two solutions observed for here for the evolution of \(m\) are also seen to occur when this problem is analyzed using the Ori model.  We discuss the implications and limitation of this approach in the Appendix.

\section{Conclusions}
\label{conclusion}

 We have studied the question of whether or not the phenomenon of mass inflation occurs in the loop black hole. In attempting to provide an answer we employed a simplified model in which the ingoing and outgoing null radiation (whose counter-streaming classically results in mass inflation) are regarded as null thin-shells. Within this framework, the generalized DTR relation allows us to make conclusions about how the metrics in the four resulting spacetime quadrants match at the intersection two-sphere. Using the DTR relation is of particular utility in our case because said matching does not directly rely on Einstein's field equation; this is good because we are attempting to work within a loop quantum gravitational framework as much as possible.

There are four  principle assumptions that were made during our analysis:

i) The LBH metric (\ref{g}) is a valid effective metric describing a black hole spacetime.

ii) The spacetime is no worse behaved on the two-sphere \(S\) than it is across the shells. Equivalently, for every point on \(S\) there is a local coordinate system in which the metric is continuous and piecewise continuously differentiable.

iii) Each spacetime quadrant bounded by the colliding shells is itself described by an effective loop metric of the form (\ref{g}).

iv) The energy of the ingoing shell as a function of \(v\) follows the same inverse power law that is known for continuous ingoing radiation.

The conclusion of our analysis is that the Cauchy horizon for the LBH remains unstable, implying that the loop black hole
is not a fully adequate effective metric for describing expected black hole corrections by loop quantum gravity. The DTR
relations force the curvature to diverge as the parameter $m_A \to -r_-/2P$, a fixed negative value dependent on the parameter
of the exterior solution. We find that $m_A$ can approach this value either directly or by passing through infinity; in  the latter situation  we have been unable to uncover any sort of curvature singularity as $m_A \to \pm\infty$.

We demonstrate in the appendix that  our results are in part corroborated by the Ori model \cite{ori}. Apart from relying  directly on Einstein's equations, this model is limited insofar as  we are unable to specify an equation of state between the pressure and energy density of the outgoing shell for the LBH.  However  in the limit the pressure is negligible  the model predicts the same behaviour that was discovered using the DTR model.

The analysis performed here will be applicable to any other quantum gravitationally corrected black holes that yield
effective spacetime metrics with inner horizons, such as found in \cite{bh2,bh3,bh4,bh5}. Indeed model calculations of the type performed in  section II for non-commutative black holes \cite{Nico} indicate that their interiors are even less stable than that of the LBH
\cite {NonCom}.  For them to avoid the same fate as the LBH will require
very special properties of the effective metric. Either they will not contain inner horizons due to their own `quantum repulsion'
(a difficult feat which is seemingly impossible for black holes with angular momenta)  or these inner horizons will have some kind of cutoff property that allows
metric functions in the B region to avoid the behaviour in (\ref{XBexp}). It will certainly not be adequate to placate the \(r=0\) singularity, the primary goal in deriving these quantum black hole solutions, to attain a purely regular spacetime.

Finally, we wish to ponder whether or not the discovery of an unstable Cauchy horizon should be an encouraging one or not. The primary goal of developing quantum gravity black hole solutions seems to be to regulate the singularities present in their classical counterparts.  Of course the structure of spacetime at the thin shell juncture and the Cauchy horizon will undoubtedly be modified by quantum gravitational effects, whatever the correct theory of quantum gravity is.  Stabilizing
the Cauchy horizon should therefore be regarded as one of the major goals of quantum gravity. However a traversable Cauchy horizon yields a spacetime (and a theory) that is no longer deterministic.
Our results indicate stabilizing the Cauchy horizon remains a major challenge for all quantum gravitational theories.

\section*{Acknowledgements}

%Thanks the Perimeter Institute for Theoretical Physics for hospitality during
%the work on this manuscript.
This research would have been impossible without the constructive discussions and technical help given by Eric Poisson. We give him the warmest of thanks for his contributions to our study.
Research at
Perimeter Institute is supported by the Government of Canada through Industry Canada
and by the Province of Ontario through the Ministry of Research \& Innovation.  This work
was supported in part by the Natural Sciences \& Engineering Research Council of Canada.

\section*{Appendix: The Ori Model}

\vspace*{0.5cm}

In this appendix we aim to outline how the Ori model of mass inflation \cite{ori} applies to the loop black hole and why it fails to be  fully predictive in this case. Even so, we will see that with a simple assumption it nevertheless gives the same result as was reached by using the DTR model.

In the Ori model  outgoing radiation is modelled as a null thin-shell, but the ingoing radiation is continuous.  In this sense
it is intermediate between the full model (in which both ingoing and outgoing streams are continuous), and the DTR model that considers \emph{both} streams as null thin-shells. We will start by considering the loop black hole in the presence of continuous ingoing radiation and how this changes the spacetime structure. Unlike the analysis in the main body of this paper, we will be using Einstein's equations to make conclusions about curvature.

\subsection{The Vaidya loop metric}

Recall that the advanced time \(v\) is defined as \(v \equiv t+r^*\), where \(r^*\) satisfies
\begin{align}
    \frac{dr^*}{dr} \equiv \frac{1}{\sqrt{G(r)F(r)}}
\end{align}
and the metric functions \(G(r)\) and \(F(r)\) are defined as in Sect. \ref{loopderiv}. If we cast the loop metric into \((v,r,\theta,\phi)\) coordinates it takes the form
\begin{align}
    ds^2=-G(r)dv^2+2\sqrt{\frac{G(r)}{F(r)}}dvdr+H(r)d\Omega^2 , 
\end{align}
where the warped function is \(H(r)=r^2+a_0^2/r^2\).
Null radiation entering a Reissner-Nordstr\(\ddot{\text{o}}\)m (RN) black hole as a continuous stream takes the form of null dust with stress-energy \(T_{\alpha \beta}=\mu(r,v)(\partial_\alpha v) (\partial_\beta v)\).   The net effect of this is to change the mass parameter from a constant to a function of \(v\): \(m \rightarrow m(v)\), with the energy density taking the form \(\mu(r,v)=m'(v)/4\pi r^2\).

For the LBH the situation is more complicated. To see this, consider the metric
 \begin{align} \label{vaidya1}
    ds^2=-G(r,v)dv^2+2\sqrt{\frac{G(r,v)}{F(r,v)}}dr dv +H(r)d\Omega^2 ,
\end{align}
where
\begin{align}
    G(r,v)&=\frac{(r-2m(v))(r-2m(v)P^2)(r+2m(v)P)^2}{r^4+a_0^2} , \nonumber \\
    F(r,v)&=\frac{(r-2m(v))(r-2m(v)P^2)r^4}{(r+2m(v)P)^2(r^4+a_0^2)} , \nonumber \\
    H(r)&=r^2+\frac{a_0^2}{r^2} ,
\end{align}
which we will refer to as the \emph{Vaidya loop black hole}.  We see that the inner and outer apparent horizons occur at \(r=2m(v)P^2\) and \(r=2m(v)\) respectively.  The Einstein equations then imply
\begin{eqnarray} 
  &&   \hspace{-0.3cm} T_{vv}^{\text{dyn}} = \frac{(1+P)^2r^2(r^4-a_0^2)(r-2m(v)P)}{4\pi(r^4+a_0^2)^2(r+2m(v)P)}m'(v) ,   \label{Tvv} \\
&&  \hspace{-0.3cm} T_{\theta \theta}^{\text{dyn}} = \frac{-P(r^4+a_0^2)}{2\pi(r+2m(v)P)^4}m'(v) \, , 
    \;\; T_{\phi \phi}^\text{dyn}=\sin^2(\theta)T_{\theta \theta}^\text{dyn} , \nonumber 
\end{eqnarray}
where the ``dyn" stands for dynamical;   we include it to emphasize that this is the stress-energy \emph{in addition} to that which is already present for the original ``static" LBH. We will call the stress-energy of the original metric, which is nonzero, \(T_{\alpha \beta}^\text{stat}\); its exact form is  unimportant.

Clearly, the stress energy in Eq. (\ref{Tvv}) does not correspond to null dust.  For \(a_0=P=0\) it reduces to the usual form of \(T_{vv}^\text{dyn}=m'(v)/4\pi r^2\). In addition,  for large \(r\) (where we know that the black hole is nearly Schwarzschild and classical theory should approximately hold) it reduces to \(T_{vv}^\text{dyn}=(1+P)^2m'(v)/4\pi r^2\), \(T_{\theta \theta}^\text{dyn}=-Pm'(v)/2\pi\) which, since the ADM mass  \(M_\text{ADM}=(1+P)^2m\), shows us that the energy component \(T_{vv}^\text{dyn}\) reduces to the usual form that we expect and that \(T_{\theta \theta}^\text{dyn}\) is of order \(P\), which is assumed small.

Ultimately, the strange form of the stress-energy results from the failure of the Einstein equation to remain a valid field equation deep inside the black hole. Note that the ingoing energy flux, which is proportional to \(T_{vv}^\text{dyn}\), becomes negative in between \(r=2m(v)P\) and \(r=\sqrt{a_0}\). If we interpret the modified stress-energy as being an effective quantum gravitational correction, then this negativity represents the correction that allows avoidance of the central singularity. Quantum gravitational effects work against the incoming radiation such that near the center of the black hole the flux becomes negative, thus avoiding collapse to a singularity. As the radiation continues to smaller \(r\) it enters the second asymptotically flat region and the flux becomes positive again, where it is then emerges as radiation from the corresponding white hole. Regardless of the interpretation, the focusing theorem guarantees  that the null energy condition must \emph{not} hold for null radiation that undergoes the ``bounce" evolution characteristic of the warped factor \(H(r)=r^2+a_0^2/r^2\). Hence the form of \(T_{\alpha \beta}^\text{dyn}\) is to be expected given the nature of the loop black hole.

A valid criticism of the preceding discussion is that, since the loop black hole was constructed as a model for the quantum gravitationally corrected vacuum solution, we should not presume to simply replace the parameter \(m\) with \(m(v)\) in the presence of ingoing null radiation.  In response, in the \(2^\text{nd}\) paragraph of Sect. \ref{massinflation}  we argued that \(m\) should be constant within a given spacetime quadrant. This reasoning is independent of how many ingoing null shells are present in the spacetime.  Conceptualizing the continuous radiation as a sequence of  ``strips",   we expect \(m\) to be constant within a given strip. The Vaidya loop metric  (\ref{vaidya1}) is the limit in which we have infinitely many ingoing null shells, each of infinitesimal energy. It follows that, given a constant \(m\) within each infinitesimally thin strip between adjacent shells, this is equivalent to \(m\) being constant along lines of constant \(v\).

In the large \(v\) limit we expect the energy density of the ingoing radiation to fall off as an inverse power, as shown by Price \cite{price}. This results from backscattered gravitational radiation, the origin of which is the surface of the collapsing star. Since this process occurs outside of the black hole, where classical theory is expected to hold to good approximation, we can be confident that the same result will apply in our case as well. This means that for large \(v\) we have
\begin{align}
    m(v)=M-\mu(v), \;\;\;\; \text{large}\;\; v , 
\end{align}
where \(M\) is the final mass of the black hole and we expect \(\mu(v) \propto v^{1-\gamma}\), where \(\gamma \geq 12\). Note that the Cauchy horizon is located at coordinates \(r=r_-\equiv 2MP^2\) and \(v=\infty\).

In addition to continuous ingoing radiation there is also a null thin-shell that acts to separate the spacetime into two regions, one to the future and one to past of the shell.  Consider the  behaviour of outgoing null geodesics in the black hole, specifically in the presence of ingoing radiation.  To this end we define a function
\begin{align} \label{f}
    f(r,v)&\equiv \sqrt{G(r,v)F(r,v)} \nonumber \\
    &=\frac{(r-2m(v))(r-2m(v)P^2)r^2}{r^4+a_0^2} . 
\end{align}
We see from Eq. (\ref{vaidya1})  that the two radial null directions are given by setting \(dv=0\) for the ingoing direction and \(2dr=fdv\) for the outgoing one. Hence
\begin{align} \label{geoeqn}
    \frac{dr}{dv}=\frac{f}{2}
\end{align}
is the equation obeyed by  the geodesic \(r(v)\)  for an outgoing photon.

Inserting \(m(v)=M-\mu(v)\) in to Eq. (\ref{f}) yields
\begin{align}
    f(r,v)=f(r,\infty)+g(r)\mu(v) , 
\end{align}
where
\begin{align}
    f(r,\infty)&=\frac{(r-2M)(r-2MP^2)r^2}{r^4+a_0^2} , \\
    g(r)&=\frac{2[(1+P^2)r-4MP^2]r^2}{r^4+a_0^2}
\end{align}
and we have neglected a term proportional to \(\mu(v)^2\) (this is justified since we are working in the \(v \rightarrow \infty\) limit). Expanding \(f(r,v)\) around the Cauchy horizon \(r_-=2MP^2\) (which is the apparent horizon in the \(v \rightarrow \infty\) limit) gives
\begin{align}
    f(r \simeq r_-,v) \simeq& f(r_-,\infty)+f'(r_-,\infty)(r-r_-) \nonumber \\
    &+g(r_-)\mu(v)+g'(r_-)(r-r_-)\mu(v) . 
\end{align}
The first term is zero (by definition) and we neglect the last term because it is 2nd order (since \(r-r_-\) is also vanishingly small). We identify
\begin{align} \label{surfgrav}
    \kappa_- = -\frac{1}{2}f'(r_-,\infty)=\frac{4M^3P^4(1-P^2)}{16M^4P^8+a_0^2}
\end{align}
as the surface gravity at the Cauchy horizon. Note that the surface gravity in the LBH is not in general given by \(\frac{1}{2}|\partial_r \sqrt{GF}|\) but rather  by
\(\kappa=\frac{1}{2}\sqrt{\frac{F}{G}}|\partial_r G|\). However at the horizons these two expressions happen to give equivalent answers. Evaluating \(g(r_-)\) in terms of \(r_-=2MP^2\) gives
\begin{align}
    g(r_-)=\frac{2(P^2-1)r_-^3}{r_-^4+a_0^2} , 
\end{align}
which is a negative quantity.

Putting this together with the null geodesic equation \(dr/dv=f/2\) yields an ODE for \(r(v)\) in the large \(v\) limit
\begin{align}
    \frac{dr}{dv}+\kappa_-(r-r_-) \simeq \frac{(P^2-1)r_-^3}{r_-^4+a_0^2}\mu(v) , 
\end{align}
whose solution is
\begin{eqnarray} 
    && \kappa_-(r-r_-)\label{soln} \\
&&    \simeq e^{-\kappa_-(v+\alpha)}+\frac{(P^2-1)r_-^3}{r_-^4+a_0^2}\sum_{n=0}^\infty \left(\frac{-1}{\kappa_-}\right)^n \frac{d^n\mu(v)}{dv^n} \nonumber ,
\end{eqnarray}
where \(\alpha\) is a constant. The summation term represents the correction to \(\kappa_-(r-r_-)\) from the \(v\) dependence; in the static case we only have the exponential term. In the \(v \rightarrow \infty\) limit we expect the exponential term to be negligible compared to the sum, and therefore will ignore it from this point on. Putting this result back into the expansion of \(f\) gives
\begin{align} \label{fsoln}
    f(r \simeq r_-,v) \simeq -\frac{(P^2-1)r_-^3}{r_-^4+a_0^2}\sum_{n=1}^\infty \left(\frac{-1}{\kappa_-}\right)^n \frac{d^n\mu(v)}{dv^n} .
\end{align}
This is the value of \(f\) that an outgoing photon will observe for large \(v\) as it travels along its geodesic.

Here we encounter an apparent mathematical contradiction. We expect that the LHS of Eq. (\ref{soln}) should be positive since \(\kappa_-\) is defined to be positive and \(r>r_-\) before the geodesic reaches the Cauchy horizon -- yet we have shown here that the RHS is negative! (For the RN black hole the result is exactly the same except that the factor in front of the sum is \(1/r_-\) , which is positive). Similarly, we see that  \(f\) is \emph{positive}, which seems to suggest that for large \(v\) the outgoing photon is no longer between the horizons -- by Eq. (\ref{geoeqn}), that its radial value \(r\) is increasing!

The problem arises in our assumption that \(r>r_-\) for large \(v\). The key difference between the RN and LBH cases is that in the presence of ingoing matter the apparent inner horizon of the RN black hole deflates, meaning that an outgoing photon has to ``chase down" the inner horizon before finally intersecting it at \(v=\infty\), where the inner apparent horizon coincides with the Cauchy horizon. In the loop black hole, however, the inner apparent horizon \emph{inflates} for increasing \(v\), as is obvious by its form \(r=2m(v)P^2\). This means that, rather than receding, the horizon instead rushes up to the meet the outgoing photon. This results in the photon first passing by the radial value that the Cauchy horizon \emph{will be at} (resulting in \(r<r_-\)) before penetrating the inflating apparent horizon (resulting in \(f>0\)), turning around and straddling the horizon before again reaching the horizon at \(v=\infty\), which becomes the Cauchy horizon.

\begin{figure}[h]
\centering
    \includegraphics[width=0.4\textwidth]{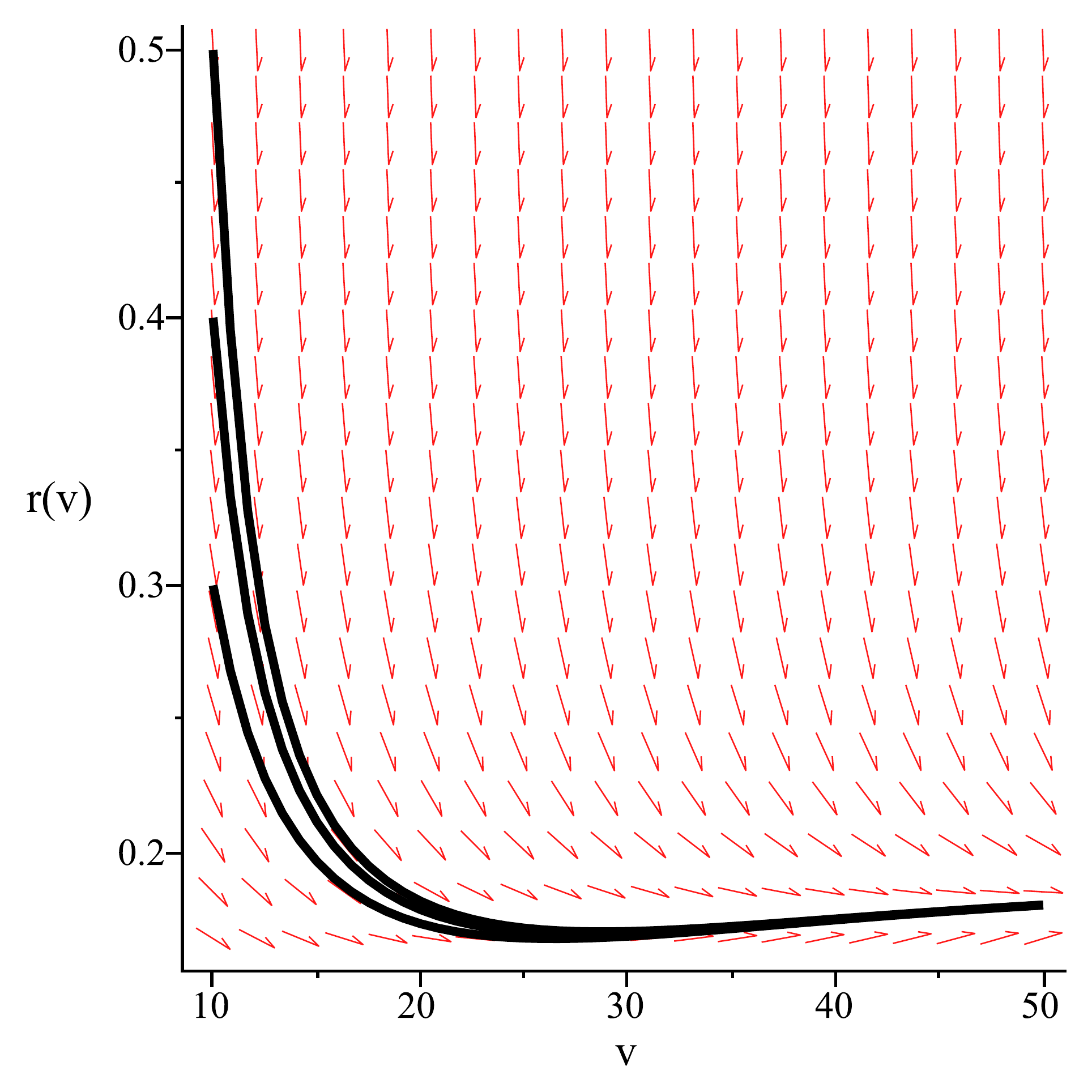}
\caption{Solution \(r(v)\) of an outgoing null geodesic, with parameters  \(M=1\), \(P^2=0.1\) and \(a_0^2=0.15\), giving  \(r_-=0.2\). The solution quickly passes by the radius \(0.2\) before coming to a stop (this is where it intersects the inner apparent horizon) and starts to increase, eventually settling onto the Cauchy horizon at \(v=\infty\). The vector field indicates the slope of \(r(v)\) at each point \((v,r)\). Specifically, this is given by \(dr/dv=f/2\).}
    \label{loopsolve}
\end{figure}

This behaviour can be seen in Fig. \ref{loopsolve}) where we have numerically solved \(r(v)\) for general \(v\) (i.e. we use the full form of Eq. (\ref{f}) in Eq. (\ref{geoeqn})) but for simplicity continued to assume an inverse power law for \(\mu(v)\). Specifically we used \(\mu(v)=(\kappa_- v)^{-1}\); this choice is fairly unphysical since the inverse power should be much higher, but this lets us see the behaviour much more easily. The same qualitative behaviour occurs for more realistic fall offs as well. The parameters used in the solution are \(M=1\), \(P^2=0.1\) and \(a_0^2=0.15\), giving the radius of the Cauchy horizon as \(r_-=0.2\). All of the stages in the evolution described above are clearly visible in the plot.
Despite this   strange and unexpected behaviour, the geodesic evolution remains perfectly well defined and the Cauchy horizon is still a surface of infinite blue-shift.

\subsection{The null thin-shell formalism}

We review here  the null thin-shell formalism \cite{thinshells,text},  outlining the results   relevant to the situation at hand. We note that the results of this section  rely directly on Einstein's equation, and so to use it within a loop quantum gravitational framework is at best a first approximation.

We wish to determine the relationship between two metrics \(g_{\alpha \beta}^\pm\) with corresponding coordinates \(x_\pm^{\alpha}\) on either side of a thin
  null shell \(\Sigma\).    Suppose  that \(\Sigma\) is equipped with some coordinate set \((\lambda,\theta^A)\), where \(\lambda\) is a parameter which runs along the null generators of the shell and \(A\) runs over the remaining two coordinates. The surface \(\Sigma\) is described by \(x_\pm^\alpha(y^a)\), with tangent vectors \(e_{\pm a}^\alpha=\partial x_\pm^\alpha/\partial y^a\); from these we identify the vector tangent to the generators \(k^\alpha\) and those transverse to them \(e_A^\alpha\)
\begin{align}
    k^\alpha \equiv e_\lambda^\alpha = \frac{\partial x^\alpha}{\partial \lambda}, \;\;\;\;\;\; e_A^\alpha=\frac{\partial x^\alpha}{\partial \theta^A} .
\end{align}
For simplicity we have removed the \(\pm\) notation; this should not introduce any ambiguity, keeping in mind that \(k\) is a vector given by \(k=\partial/\partial \lambda\), and the only thing that changes between the two sides of the shell is the coordinates in which we evaluate this vector.

Since this is a null shell, we have
\begin{align}
    k_\alpha k^\alpha = 0 = k_\alpha e_A^\alpha
\end{align}
and the induced metric is given by
\begin{align}
    \sigma_{A B} \equiv g_{\alpha \beta}e_A^\alpha e_B^\beta .
\end{align}
In order that the shell itself have a well-defined intrinsic geometry the induced metric should be independent of which side of the shell it is being evaluated on, implying
\begin{align} \label{metcont}
    [\sigma_{AB}]=0 .
\end{align}

To complete the basis over all spacetime dimensions we need to include another vector that isolates the part of the metric \(g_{\alpha \beta}\)  transverse to the shell. To this end, we introduce an auxiliary vector \(N^\alpha\) that solves
\begin{align} \label{auxiliary}
    N_\alpha N^\alpha=0=N_\alpha e_A^\alpha, \;\;\;\;\;\; N_\alpha k^\alpha=-1 .
\end{align}
These conditions specify \(N^\alpha\) completely. Note, however, that one needs to compute this vector separately for each side of the shell. With this, the total metric can be shown to be given by the completeness relation
\begin{align}
    g^{\alpha \beta}=-k^\alpha N^\beta-N^\alpha k^\beta+\sigma^{AB}e_A^\alpha e_B^\beta ,
\end{align}
where \(\sigma^{AB}\) is the matrix inverse of \(\sigma_{AB}\) (note that this is just a special case of Eq. (\ref{completeness})).

We also introduce what is known as the \emph{transverse curvature}  \(C_{ab}=\frac{1}{2}(\mathcal{L}_N g_{\alpha \beta})e_a^\alpha e_b^\beta\), which more explicitly is
\begin{align}
    C_{ab}=-N_\alpha e^\alpha_{a\, ; \beta}e^\beta_b , 
\end{align}
where Latin indices run over three coordinates: \(y^a=(\lambda,\theta^A)\).
The stress-energy of the shell is \cite{text}
\begin{align}
    T^{\alpha \beta}_\Sigma=(-k_\mu u^\mu)^{-1}S^{\alpha \beta}\delta(\tau) ,
\end{align}
where \(u^\mu\) is tangent to an arbitrary congruence of timelike geodesics parameterized by \(\tau\) in which the geodesics cross \(\Sigma\) at \(\tau=0\). The \(\delta(\tau)\) factor represents the assumption that the shell is infinitely thin, and is equivalent to the divergence of the Einstein tensor mentioned above. The tensor \(S^{\alpha \beta}\) is given by
\begin{align}
    S^{\alpha \beta}=\mu k^\alpha k^\beta+j^A(k^\alpha e_A^\beta+e_A^\alpha k^\beta)+p \sigma^{AB}e_A^\alpha e_B^\beta , 
\end{align}
where the energy density, current and surface pressure are given by
\begin{align} \label{bareSE}
    \mu=-\frac{1}{8\pi}\sigma^{AB}[C_{AB}],& \;\;\; j^A=\frac{1}{8\pi}\sigma^{AB}[C_{\lambda B}], \nonumber \\
    p=-&\frac{1}{8\pi}[C_{\lambda \lambda}] . 
\end{align}
Note that the factor \((-k_\mu u^\mu)^{-1}\) means that \(T^{\alpha \beta}_\Sigma\) is observer dependent; a measurement of the shell's stress-energy will depend on the geodesic followed by the observer doing the measuring. We can thus define the ``physical" quantities measured by this observer:
\begin{align} \label{press1}
    \mu_{\text{phys}}=(-k_\mu u^\mu)^{-1}\mu,& \;\;\; j^A_{\text{phys}}=(-k_\mu u^\mu)^{-1}j^A, \nonumber \\
    p_{\text{phys}}&=(-k_\mu u^\mu)^{-1}p .
\end{align}
Noting that \(C_{\lambda \lambda}=-N_\alpha k^\alpha_{\; ;\beta}k^\beta=\kappa\), where   \(\kappa\) is the ``acceleration" of the shell, \(k^\alpha_{\; ;\beta}k^\beta=\kappa k^\alpha\), we obtain from Eq. (\ref{bareSE}) the matching condition
\begin{align} \label{match1}
    [\kappa]=-8\pi p ,
\end{align}
which, via Raychaudhuri's equation (and making use of hypersurface orthogonality),   is equivalent to
\begin{align}  \label{match2}
    [\kappa]\theta=8 \pi [T_{\alpha \beta}k^\alpha k^\beta] , 
\end{align}
with \(\theta\)   the expansion of \(\Sigma\) (with respect to the generating parameter \(\lambda\)).  \(T_{\alpha \beta}\) is the stress-energy of the surrounding spacetime, \emph{not} of the shell.

We close this subsection by considering what happens when we choose a different generating parameter, \(\bar{\lambda}\), to describe the shell. Under the reparametrization 
\begin{align}
    \lambda \rightarrow \bar{\lambda}(\lambda,\theta^A) ,
\end{align}
we define
\begin{align}
    g\equiv \frac{\partial \bar{\lambda}}{\partial \lambda}, \;\;\;\;\;\; c_A \equiv \frac{\partial \bar{\lambda}}{\partial \theta^A} , 
\end{align}
implying that the generating vector transforms as
\begin{align}
    \bar{k}^\alpha=g^{-1}k^\alpha
\end{align}
and,  less obviously, the stress-energy quantities defined in Eq. (\ref{bareSE}) transform as
\begin{eqnarray}
   && \bar{\mu} = g\mu+2c_Aj^A+c^Ac_Ag^{-1}p, \nonumber \\
  &&  \bar{j}^A = j^A+c^Ag^{-1}p, \nonumber \\
   && \bar{p} =g^{-1}p .  \label{press2}
\end{eqnarray}
In total it can be shown that the stress-energy transforms as \(\bar{S}^{\alpha \beta}=g^{-1}S^{\alpha \beta}\) and therefore that the combination \((-k_\mu u^\mu)^{-1}S^{\alpha \beta}\) is invariant under the the reparameterization. That is, we have invariance of the total stress-energy of the shell \(T^{\alpha \beta}_\Sigma\).

 \subsection{The evolution of \(m_+(v)\)}

\begin{figure}[h]
  \centering
 %\begin{center}
    \includegraphics[width=0.45\textwidth]{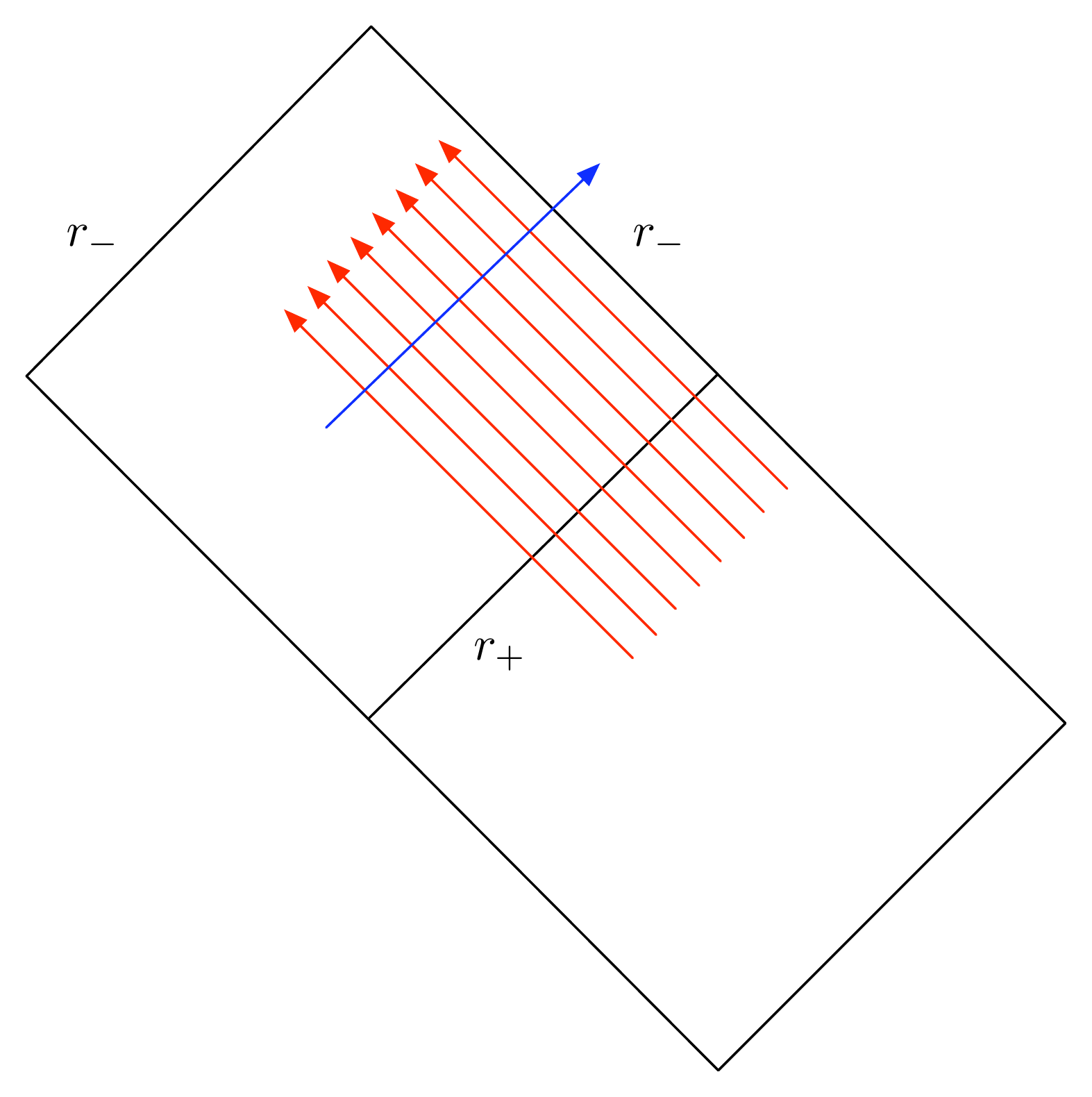}
%    \end{center}
  \caption{Schematic illustrating the situation considered in the Ori model. We see both the continuous ingoing radiation as well as the outgoing shell \(\Sigma\). This shell acts to split the spacetime into two regions.%\textcolor{red}{[again, we should come up with a better figure before publishing]}
  }
    \label{oripic}
\end{figure}

Taking  ingoing radiation to be continuous, giving us the Vaidya spacetime Eq. (\ref{vaidya1}), and outgoing radiation   modeled as a null thin-shell \(\Sigma\) that acts to split the spacetime into a future region (+) and a past region (-),  gives
a  scenario   displayed in Fig. \ref{oripic}.

Since on either side of the shell only an influx of radiation is present, it follows that the spacetime should be described by a Vaidya loop metric on both sides, given by
\begin{align} \label{vaidya2}
    ds_\pm^2=-G_\pm dv_\pm^2+2\sqrt{\frac{G_\pm}{F_\pm}}dr dv_\pm +Hd\Omega^2 ,
\end{align}
where the metric functions \(G_\pm\) and \(F_\pm\) are defined as above, but with \(m(v)\) replaced by \(m_\pm(v_\pm)\).  Henceforth   for convenience we will drop the subscript on the advanced time on the past side \(v_- \equiv v\) since we expect that here the spacetime behaves as outlined in part A of this appendix. The mass will be given by the function \(m_-(v)=M-\mu(v)\) and the coordinate \(v\) will still determine the location of the Cauchy horizon (at \(r=2MP^2\) and \(v=\infty\)).

To investigate the behaviour of \(m_+\) , we begin by noting that the tangent vectors to the shell  transverse to the generators are
\begin{align}
    e^\alpha_\theta=\frac{\partial x^\alpha}{\partial \theta}=(0,0,1,0), \;\;\;\;\; e^\alpha_\phi=\frac{\partial x^\alpha}{\partial \phi}=(0,0,0,1) .
\end{align}
The induced metric is thus
\begin{align}
    \sigma_{AB}=H(r)
    \begin{pmatrix}
    1 & 0 \\
    0 & \sin^2\theta
    \end{pmatrix} ,
\end{align}
from which the condition specified by Eq. (\ref{metcont}), \([\sigma_{AB}]=0\), gives us \([H(r)]=0\). The solution that we take from this condition is that the coordinate \(r\) is continuous across the shell.

With the knowledge that \(r\) is continuous across \(\Sigma\) we can immediately read off the geodesic of the shell as seen from either the past or future side (i.e. in either coordinate system), and in so doing also obtain a relation between \(v\) and \(v_+\) (which is valid \emph{only} along \(\Sigma\))
\begin{align}
    2dr=f_+dv_+=f_-dv, \;\;\;\;\;\;\; \text{along} \; \Sigma ,
\end{align}
where for convenience we have defined
\begin{align}
    \hspace{-0.3cm} f_\pm=\sqrt{G_\pm F_\pm}=\frac{(r-2m_\pm(v))(r-2m_\pm(v)P^2)r^2}{r^4+a_0^2} .
\end{align}
It it important to keep in mind that in all subsequent equations the coordinate \(r\) is actually a function of \(v\) since everything is evaluated on the shell. Specifically, the form of \(r=r(v)\) is given by Eq. (\ref{soln}) and looks like the solutions plotted in Fig. \ref{loopsolve}.

Defining the tangent vector \(k^\alpha\) to be
\begin{align}
    k^\alpha=\frac{\partial x^\alpha}{\partial v} , 
\end{align}
we find
\begin{align}
    k_-^\alpha=\frac{\partial x_-^\alpha}{\partial v}=(1,\frac{f_-}{2},0,0), \;\; k_+^\alpha=\frac{\partial x_+^\alpha}{\partial v}=(\frac{f_-}{f_+},\frac{f_-}{2},0,0) ,
\end{align}
 on the past and future side of shell, where
 \(x_\pm^\alpha=(v_\pm,r,\theta,\phi)\) and \(v_- \equiv v\).
We wish to compute the acceleration \(\kappa=-N_\alpha k^\alpha_{\; ;\beta}k^\beta\) on either side so that we can use the equation \([\kappa]=-8\pi p\). In order to do this we need to first compute the auxiliary vector \(N^\alpha\); the form of which is fully determined by the conditions in Eq. (\ref{auxiliary}). Unlike \(k\), \(N^\alpha\)   differs   on the two different sides of the shell.
  On the past side the conditions on \(N^\alpha\) yield
\begin{align}
    N_-^\alpha=(0,-\sqrt{\frac{F_-}{G_-}},0,0)
\end{align}
and on the future side they give us
\begin{align}
    N_+^\alpha=(0,-\frac{F_+}{f_-},0,0) .
\end{align}

Computing  \(\kappa=-N_\alpha k^\alpha_{\; ;\beta}k^\beta\) on both sides of \(\Sigma\)
and noting that \(m_\pm'(v_+)=\frac{f_+}{f_-}m_\pm'(v)\) yields the differential equation
\begin{align} \label{ODE}
    &\frac{r^3(1+P)^2(r-2m_+P)}{(r^4+a_0^2)(r+2m_+P)}\frac{1}{f_+}\frac{dm_+}{dv} \nonumber \\
    -&\frac{r^3(1+P)^2(r-2m_-P)}{(r^4+a_0^2)(r+2m_-P)}\frac{1}{f_-}\frac{dm_-}{dv}  \nonumber \\
    +&\frac{P}{(r+2m_-P)(r+2m_+P)}f_-(m_--m_+)=-4\pi p , 
\end{align}
where \(m_-=m_-(v)=M-\mu(v)\) is a known function and \(r=r(v)\) is the solution of \(r'(v)=f_-/2\) (as per Eq. (\ref{soln})). From this  the metric functions are purely functions of \(v\): \(f_\pm=f_\pm(r(v),m_\pm(v))\). This is because we are evaluating everything on the shell, which follows a known geodesic.

Note that we could just as easily have done this analysis using \(r\) to parameterize \(\Sigma\) instead of \(v\).  Replacing \(v \rightarrow r\) then \(p \rightarrow (\partial v/\partial r)p=(2/f_-)p\) which, when substituted into   Eq. (\ref{ODE} yields the same
equation  but with the RHS multiplied by \(f_-/2\).

 \begin{figure}[h!]
  \centering
  \subfloat{\label{mp2}\includegraphics[width=0.4\textwidth]{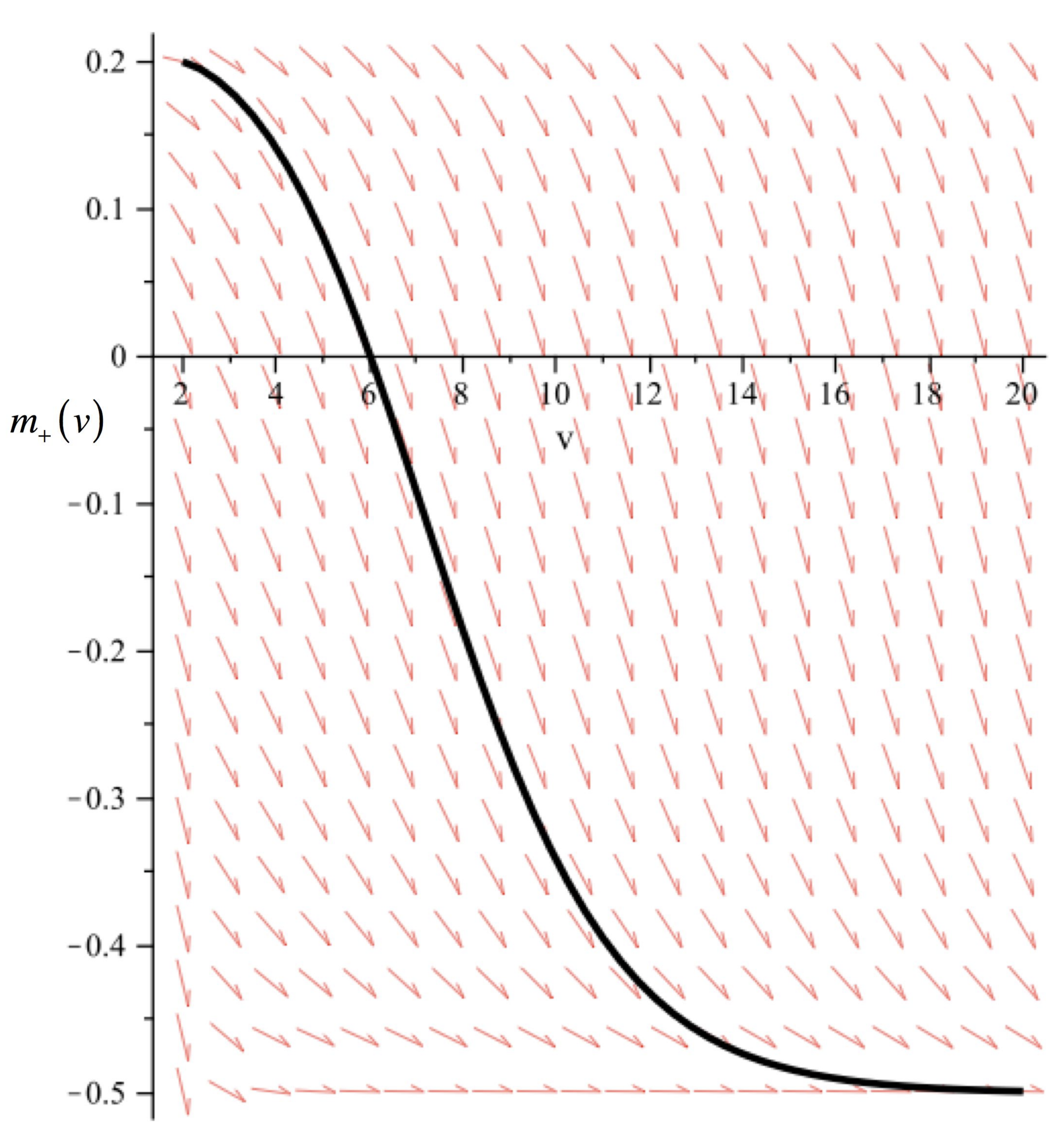}}
  \;
  \subfloat{\label{mp1}\includegraphics[width=0.4\textwidth]{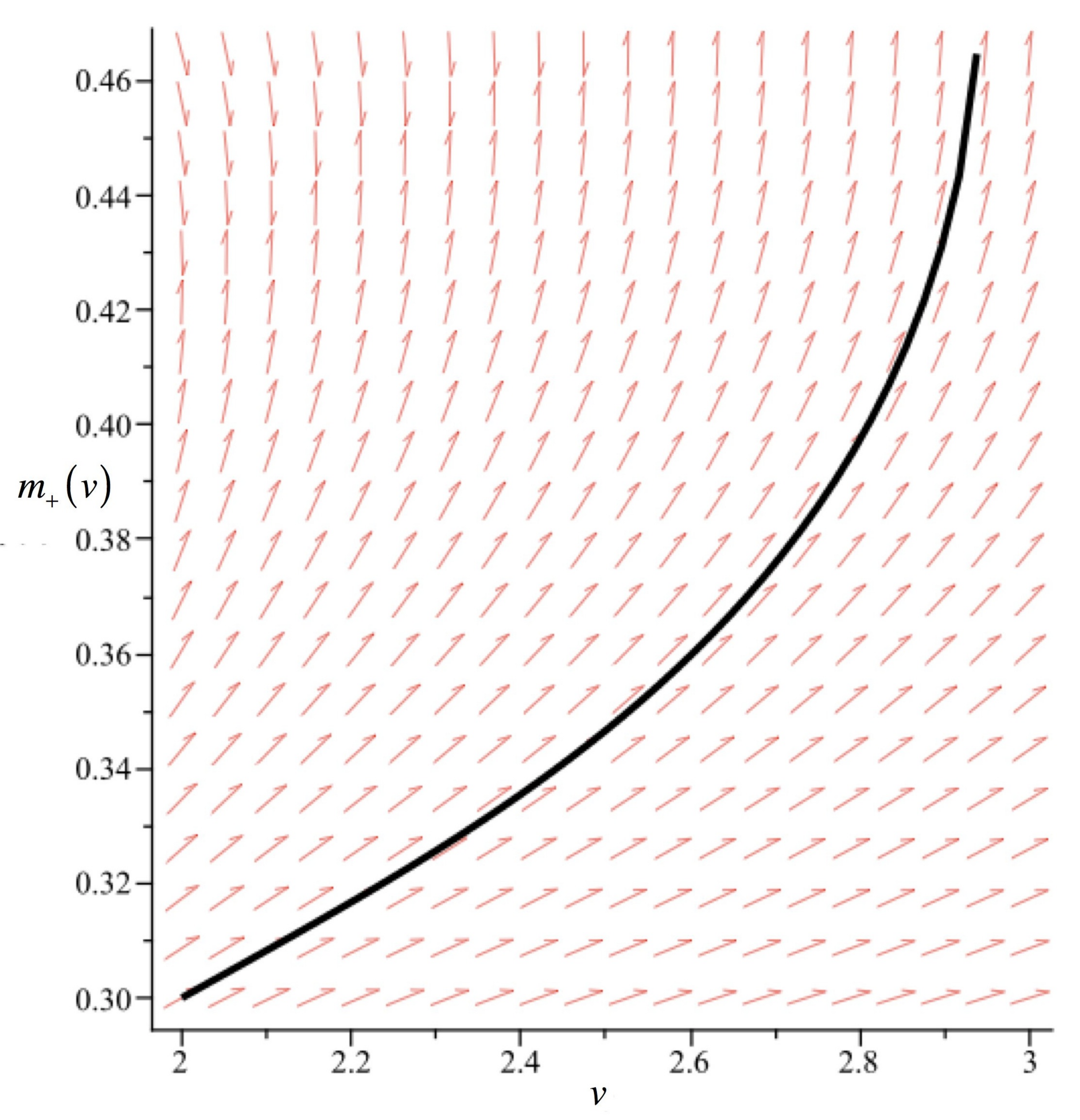}}
  \caption{The results of numerically integrating Eq. (\ref{ODE}) with parameters \(M=1\), \(P^2=0.25\) and \(a_0^2=0.1\); this gives \(-r_-/2P=-MP=-0.5\). For the top graph the initial conditions were \(m_+=0.2\), which results in \(m_+(v)\), asymptotically approaching \(-r_-/2P=-0.5\). For  the bottom graph the initial conditions were \(m_+=0.3\), which results in \(m_+(v)\) diverging to infinity at finite \(v\). These are the two solutions obtained from the DTR analysis.}
  \label{mp}
\end{figure}

The only thing left to do before examining the behaviour of \(m_+\) is to specify the pressure \(p\). Unlike the RN case
we cannot assume that the shell is pressureless due to Eq. (\ref{Tvv}).   We therefore need
some other equation of state (EoS)  \(p=p(\mu)\), where \(\mu\) is the energy-density of the shell, to specify the pressure.  Using Eq. (\ref{bareSE}) we can compute the energy density of the shell \(\mu=-\frac{1}{8\pi}\sigma^{AB}[C_{AB}]\),yielding us some function of \(r\), \(m_-\) and \(m_+\) which, given \(p=p(\mu)\), can be inserted
into the RHS of Eq. (\ref{ODE}) to give us an equation without any unknown, free parameters.

There are two equivalent ways in which we could try to obtain an EoS given our model. The first is to look at the stress-energy of the ingoing radiation, Eq. (\ref{Tvv}), and attempt to simply read off the appropriate EoS. As can quickly be seen, however, this simply is not possible given the form of the stress-energy we have; an EoS does not exist. Alternatively, we could compute the energy density \(\mu\) and pressure \(p\) of the shell \(\Sigma\) (using Eq. (\ref{bareSE})) in the case that there is no ingoing radiation.
 For example, if we compute the pressure in this scenario we obtain Eq. (\ref{ODE}) but with the first two terms set to zero.  However
 this entails the same problems as the first approach, because the scenario of the outgoing shell in an otherwise static black hole is really just a special case of an outgoing Vaidya spacetime, and so the stress-energy of these two situations will be of the same form.

  Without being able to specify the pressure of the shell, the Ori model for the LBH loses its predictive power; each choice for the  \(p\) will   give a different evolution for \(m_+(v)\).  Despite this, it is interesting to note that  for the choice \(p=0\) the solution \(m_+(v)\) of Eq. (\ref{ODE}) displays the same behaviour as that   obtained for \(m_A(v)\) in the DTR analysis. That is, we find that if the initial condition for \(m_+\) is small enough then it evolves continuously to a value of \(-r_-/2P\), and if the initial condition is higher then the solution is seen to diverge to infinity for finite \(v\). These are exactly the two options found for \(m_A\).

To see this explicitly, we numerically solved Eq. (\ref{ODE}) in the large \(v\) limit. We set the parameters to \(M=1\), \(P^2=0.25\) and \(a_0^2=0.1\); this gives \(-r_-/2P=-MP=-0.5\). The power law decay was set to \(\mu(v)=(\kappa_- v)^{-3}\) and the functions \(r(v)\) and \(f_-(v)\) that we used are given by Eqs. (\ref{soln}) and (\ref{fsoln}) respectively, where we retained the first two terms in the sums. Fig. \ref{mp}) displays the result of numerically integrating the ODE, where in the top figure we set the initial condition \(m_+\) to \(0.2\) and in the bottom figure it was set to \(0.3\). The fact that these solutions match the DTR results is quite intriguing, and suggests that perhaps using Einstein's classical field equation in this instance is a more reliable approximation than might originally be expected.

\end{document}